\documentclass{ptephy_v1}

\preprintnumber{XXXX-XXXX} 

\usepackage{CJKutf8, color}

\usepackage{graphicx}
\usepackage{bm,url,braket,color}

\usepackage{ulem}
\DeclareRobustCommand{\erase}{\bgroup\markoverwith{\textcolor{black}{\rule[.5ex]{2pt}{0.4pt}}}\ULon}

\renewcommand{\thesubsection}{\Alph{subsection}}

\begin{document}

\title{Cooling of Hybrid Stars with a 2SC+$\braket{dd}$ Phase}

\author[1]{Tsuneo Noda\footnote{noda@kurume-it.ac.jp}}
\author[2]{Akira Dohi
\footnote{akira.dohi@riken.jp}}
\author[3,4]{Nobutoshi Yasutake}
\author[5]{Huan Chen}
\author[4]{Toshiki Maruyama}
\author[6]{Toshitaka Tatsumi}

\affil[1]{Department of Education and Creation Engineering, Kurume Institute of Technology, 2228-66 Kamitsu-machi, Kurume, Fukuoka 830-0052, Japan}
\affil[2]{Interdisciplinary Theoretical and Mathematical Sciences (iTHEMS), RIKEN, 2-1 Hirosawa, Wako, Saitama 351-0198, Japan}
\affil[3]{Physics Department, Chiba Institute of Technology, 2-1-1 Shibazono, Narashino, Chiba 275-0023, Japan}
\affil[4]{Advanced Science Research Center, Japan Atomic Energy Agency, 2-4 Shirakata, Tokai, Ibaraki 319-1195, Japan}
\affil[5]{School of Mathematics and Physics, China University of Geosiciences, Lumo Road 388, 430074 Wuhan, China}
\affil[6]{Kitashirakawa-Kamiikeda-Cho,Kyoto 606-8287,Japan}

\begin{abstract}%
In 2019, Fujimoto, Fukushima, \& Weise have proposed a new color-superconductive state, 2SC+$\braket{dd}$ phase, which can be smoothly connected to the low-density baryon superfluidity in contrast to the 2SC phase. In this scenario, the neutron ${}^3P_2$ superfluidity on the low-density side of the phase transition is inherited by unpaired $d$-quarks in the 2SC phase on the high-density side.
Since this could be realized in hybrid stars (neutron stars containing hadronic and quark matter), the 2SC+$\braket{dd}$ phase may change the properties of neutron stars compared to the traditional 2SC phase. In this work, we study the thermal evolution of hybrid stars with the 2SC+$\braket{dd}$ phase for the first time. We find that NSs with the 2SC+$\braket{dd}$ phase become hotter than those with the 2SC phase, and are close to the CFL phase. The ${}^{3}P_2$ superfluidity plays an important role in cooling curves with not the 2SC but 2SC+$\braket{dd}$ phases due to the suppression of quark $\beta$ decay. We therefore point out that, if the scenario of 2SC+$\braket{dd}$ phase is true, it could be specified through low-temperature observations such as Vela, 3C58, Vela Jr., and Vela-like pulsar.
\end{abstract}

\subjectindex{D41,E32}

\maketitle

\newcommand{\apj}{Astrophys. J. }
\newcommand{\apjs}{Astrophys. J. Suppl. }
\newcommand{\apjl}{Astrophys. J. Lett. }
\newcommand{\pasj}{Publ. Astron. Soc. Japan. }
\newcommand{\pasa}{Publ. Astron. Soc. Australia. }
\newcommand{\physrep}{Phys. Rep. }
\newcommand{\ptp}{Prog. Theor. Phys. }
\newcommand{\ptps}{Prog. Theor. Phys. Suppl. }
\newcommand{\ptep}{Prog. Theor. Exp. Phys. }
\newcommand{\AIP}{AIP Conf. Proc. }
\newcommand{\aap}{Astron. Astrophys. }
\newcommand{\ssr}{Space Sci. Rev. }
\newcommand{\nat}{Nature }
\newcommand{\sci}{Science }
\newcommand{\prc}{Phys. Rev. C}
\newcommand{\prd}{Phys. Rev. D}
\newcommand{\prl}{Phys. Rev. Lett. }
\newcommand{\nar}{New Astron. Rev. }

\newcommand{\araa}{Ann. Rev. Astron. Astrophy. }
\newcommand{\mnras}{Mon. Not. Roy. Astron. Soc. }

\newcommand{\nphysa}{Nucl. Phys. A}

\newcommand{\jcap}{JCAP}

\newcommand{\memsai}{Memorie della Soc. Astron. Ital.}

\section{Introduction}

The densest star in the universe, neutron star (NS), has supra nuclear saturation density $\rho_{\rm nuc}\simeq2.8\times10^{14}~{\rm g~cm}^{-3}$ in their centres. In such extremely high-density matter, all baryons such as neutrons, protons, and hyperons will eventually melt and change into unbound quarks and gluons. The behavior of such a quark-hadron phase transition is still uncertain
but in ultra high-density regions, deconfined quarks and gluons will compose matter since the quark-quark interactions become very weak due to the asymptotic freedom. Quark-hadron phase transition must therefore occur in the \textit{intermediate}-density regions, where the NS core is realized. Therefore, elucidating the state of quark matter and hadron matter in the \textit{intermediate}-density region has been a challenge in understanding the structure of neutron star cores.

In quark matter, the colour superconductive (CSC) phase appears in cold NSs~(e.g, Ref. \cite{2011RPPh...74a4001F}). Roughly speaking, there are mainly two kinds of CSCs. One is the colour flavour-locked (CFL) phase, where all deconfined quarks are paired, while the other is the 2-flavour colour superconducting (2SC) phase, where only all strange quarks and all quarks with a colour (blue is usually chosen~\cite{2002PhRvD..66i4007S,2003PhRvD..67e4018A}) are ungapped~\citep{1984PhR...107..325B}. Because the CFL phase is the ground state in the quark matter when quark chemical potential is large enough to exceed masses of all quarks (especially strange quark), it is likely to appear in very high-density regions according to the perturbation QCD method~\cite{1998PhLB..422..247A,1999NuPhB.537..443A}.
Compared with the CFL phase, the 2SC phase may occur in lower-density regions\footnote{The possibility of 2SC in compact stars is still under discussion. Some studies (cf. \cite{2002JHEP...06..031A}) indicate that the 2SC phase is energetically unfavourable in compact stars, but the others (cf. \cite{2006NuPhA.768..118A}) show that the 2SC phase can appear in compact stars with a strong coupling.}~\cite{2002PhRvD..66i4007S,2003NuPhA.714..481N,2008RvMP...80.1455A}.

In hadronic matter, neutron superfluidity in the ${}^{3}P_2$ state (and proton superconductivity in the ${}^{1}S_0$ state) are believed to develop over saturation density. The most promising reason for realising ${}^{3}P_2$ superfluidity is the measurement of nucleon-nucleon scattering phase shifts~\citep{1970PThPh..44..905T,2013arXiv1302.6626P}. It is shown that the phase shift is higher in the ${}^{3}P_2$ superfluidity than ${}^{1}S_0$, implying strong neutrons ${}^{3}P_2$ pairing in the hadronic core, although its density dependence is still uncertain. Regarding protons, the ${}^{1}S_0$ superconductivity is likely to appear owing to their small fraction at most $\lesssim20\%$, which effectively lowers bulk density~\citep{1993PThPS.112...27T,1993NuPhA.555..128W}. Thus, to reveal the quark-hadron phase transition, the continuity between neutrons ${}^{3}P_2$ pairing and $ud$ quarks 2SC pairing should be important~\footnote{The continuity between neutrons ${}^{3}P_2$ pairing and CFL pairing could also be considered, but because of large strange quark mass, $SU(3)$ flavour symmetry is broken in the density regions where quark chemical potential is not dominant. As a consequence, $ud$ quarks pairing is natural to be considered in the intermediate density.}.

Although the conventional picture of quark-hadron phase transition is the first-order phase transition between hadronic and quark matter in the pure 2SC phase, they would not be smoothly connected because the chiral symmetry is broken in the former, but restored in the latter. As the solution, Fujimoto, Fukushima, \& Weise~\citep{2021PhRvD.104f3036K} pointed out that ${}^{3}P_2$ neutrons superfluidity may be transformed to the ${}^{3}P_2$ $d$-quark one, which coexists with paired $ud$ quarks (so called a 2SC+$\braket{dd}$ phase), from symmetry consideration. That is, chiral symmetry and baryon number conservation are permitted to break spontaneously and simultaneously near the hadron-quark transition density, resulting in the emergence of the quark CSC phase and the baryon superfluid phase respectively.
As a result, the $\braket{dd}$ condensate could be a key for quark-hadron continuity since it can smoothly connect with the baryon superfluid phase as well as the quark CSC phase due to common symmetry. 

Physics of ultrahigh dense matter can be proven by the observation of NSs. There are mainly two approaches for this purpose. One is the overall structure of NSs, i.e., the $M$--$R$ relation. The observation of masses of NSs is the strong constraint on the equation of state (EOS), such as the observations of $2M_\odot$ NSs~\citep{2010Natur.467.1081D,2013Sci...340..448A,2020NatAs...4...72C,2022ApJ...934L..17R}. The radii observations are also important such as the recent gravitational wave observations and X-ray telescope observations (e.g., for reviews of observational constraints, see Refs. \cite{2021PrPNP.12003879B,2021Univ....7..182L}). Using the approach with $M$--$R$ relation, the \textit{bulk} properties of NS matter are restricted. The other is to determine the state of high-dense matter. For this purpose, isolated NSs are one of the good observational targets to fix the cooling curves~(for reviews, see Refs. \cite{2004ARA&A..42..169Y,2006NuPhA.777..497P,2020IJMPE..2930007K,2021PrPNP.12003879B}). There are important observations that play the role of constraint on NS matter; Cassiopeia A observation illuminates the importance of ${}^3P_2$ superfluidity~\citep{2011PhRvL.106h1101P,2011MNRAS.412L.108S}; 3C58 and Vela highlight the requirement of a strong neutrino emission process with mild suppression by superfluidity.

Considering the quark-hadron phase transition, there are pairing effects on the cooling process to satisfy the cooling observations. Too strong neutrino emission such as quark $\beta$-decay or hadronic direct Urca (DU), should be moderately suppressed by adjusting the gap. This is possible in hadronic matter due to the appearance of ${}^3P_2$ superfluidity of neutrons, but not possible in the case of the 2SC phase in quark matter. The impact of such CSC phases on cooling curves has been well examined~\citep{2000PhRvL..85.2048P,2000ApJ...533..406B,2001A&A...368..561B,2005PhRvC..71d5801G,2013ApJ...765....1N,2016EPJA...52...44S}. In particular, Grigorian et al. (2005)~\cite{2005PhRvC..71d5801G} calculated cooling curves of hybrid stars with the 2SC phase and an additional species ``X", named as 2SC+X phase, where the density-dependent gap $\Delta_{X}$ is introduced as an ansatz~\footnote{Grigorian et al. (2005) \cite{2005PhRvC..71d5801G} assumes the density dependence of neutrons superfluidity (SF) in the ${}^{3}P_2$ state~\citep{2004PThPh.112...37T}. Hence, the $X$ phase in their case would be apart from ${}^{3}P_2$ superfluidity.}. They showed that the compatible gap energy of the ``X" phase with the cooling data is $\Delta_{X}\sim30~{\rm keV}$, which is surprisingly similar to the typical ${}^{3}P_2$ neutrons superfluid gap $\Delta_{nn}\sim\mathcal{O}(0.01-0.1)~{\rm MeV}$~(for a review, see Refs.~\cite{2013arXiv1302.6626P,2019EPJA...55..167S}), as pointed by Ref.~\cite{2020PhRvD.101i4009F}. In that sense, the scenario of 2SC+$\braket{dd}$ seems observationally favorable~\cite{2022EPJWC.26011024N}. Considering 2SC+$\braket{dd}$ phase, it is not necessary to consider the free parameter about the ``X'' which comes from completely mysterious origin. Also, in the 2SC+X scenario, ${}^3P_2$ and ``X'' coexist and it does not match with the context of Ref. \cite{2020PhRvD.101i4009F}.
In this work, we investigate the thermal evolution of hybrid stars in treating the ${}^{3}P_2$ states associated with both neutrons and $\braket{dd}$ quark without introducing the hypothetical X phase. In particular, we mostly focus on the critical temperature of ${}^{3}P_2$ superfluidity, which is highly uncertain~(e.g., Ref.~\cite{2004PThPh.112...37T}). 

This paper is organized as follows: In Section
\ref{sec:models}, we present the EOS adopted in this study. After introducing features of our EOS, our treatment of the neutrino cooling processes, baryon superfluidity, and quark CSC are explained, as well as the setup of our cooling calculation. Section \ref{sec:results} shows cooling curves and temperature structure with the 2SC+$\braket{dd}$ phase, and investigates the dependence of ${}^{3}P_2$ superfluidity critical temperature. 
Section  \ref{sec:conc} devotes the conclusion. 

\section{Cooling Models}
\label{sec:models}

\subsection{EOS}
\label{subsec:eos}

We construct the hybrid EOS with the first-order phase transition, based on the method described in Ref.~\cite{2016JPhCS.665a2068Y}: In the hadronic phase, the Brueckner-Hartree-Fock method is utilized with the use of realistic two-body nucleon-nucleon potential (Bonn-B (BOB) potential \cite{1987PhR...149....1M}) and phenomenological three-body potential (Urbana UIX \cite{1997PhRvC..56.1720P}). To investigate the concept of quark-hadron continuity, a crossover equation of state (EOS) is required in principle. However, it is difficult to determine the particle number fractions in the crossover region, as well as physical quantities such as neutrino emissivity, which are essential for cooling calculations. These values can be determined using a first-order phase transition equation of state. For simplicity, we calculate stellar structure using a first-order phase transition equation of state and assume that quark-hadron continuity holds under the particle fractions determined by the first-order phase transition model.

Neutrons and protons with their masses of $m_n =m_p = 939~{\rm MeV}$ are contained, but the degree of hyperons is not considered. Due to the uncertainties of hyperon interactions~(see Ref. \cite{2020PrPNP.11203770T} for a review), the hyperons may play a role in the EOS. In the case of hybrid matter, however, the effects of hyperon may be relatively small because the onset density of $\Lambda$ hyperons may be generally similar (or a little lower) to that of the quark phase in case of first-order phase transition~\citep{2007PhRvD..76l3015M,2015ApJ...813..135M,2018PhRvD..97b3018B,2023PhRvD.107j3009Q}. Thus, the qualitative effects to soften the EOS can be similarly treated for both hyperons matter and quark matter. Regarding the rapid $\nu$ cooling processes, hyperon Urca processes~\citep{1992ApJ...390L..77P} must work in particular for $\Lambda$-induced Urca~\cite{2006PThPh.115..355T,2009ApJ...691.1035H} but their dominant density regions would be very narrow due to the presence of quark $\beta$ decay. Thus, the absence of a degree of hyperons could be well justified in our EOS. 

In the quark phase, we employ the Dyson-Schwinger approach following Ref.~\cite{2011PhRvD..84j5023C}; To obtain the quark number density, which gives the all information of the EOS, we need to obtain the quark propagator $S(p;\mu)$ with momentum $p$ at finite quark chemical potential $\mu$, that is, solve the QCD's Dyson-Schwinger equation~\citep{1994PrPNP..33..477R}\footnote{We use the standard notation and conventions of Minkowski space formulation, such as ${\slash \hspace{-0.22cm}}{A}=A_{\mu}\gamma^{\mu}=g^{\mu\nu}A_{\mu}\gamma_{\nu}, \left\{\gamma^{\mu},\gamma^{\nu}\right\}=2g^{\mu\nu}, \gamma_\mu^{\dagger}=\gamma_\mu$ and $\gamma_5=i\gamma^0\gamma^1\gamma^2\gamma^3\gamma_4$, where $g_{00}=1$ and $g_{\mu\mu}=-1$ for $\mu\neq0$. 
}.

\begin{eqnarray}
S^{-1}(p,\mu) = Z_2\left[i{\slash \hspace{-0.22cm}}{p}-\gamma_4\mu+m_q\right]+\Sigma(p;\mu)~,
\label{eq:2a.1}
\end{eqnarray}
where $m_q$ is the current quark mass, where we set $m_u=m_d=0$ and $m_s=115~{\rm MeV}$. $\Sigma$ denotes the normalized self-energy. However, it is highly uncertain since it includes the non-perturbative terms such as the gluon propagator and the quark-gluon vertex at finite $\mu$ regions. Since this formalism must be connected with the hadronic physics at $\mu=0$,  we assume following self-energy extended from that in $\mu=0$ as
\begin{eqnarray}
\hspace*{-1cm}
    \Sigma(p;\mu) &\approx& Z_1\int^{\Lambda_{\rm UV}}\frac{d^4q}{\left(2\pi\right)^4}{\mathcal G}\left(k^2;\mu\right)D_{\rho\sigma}(k) \gamma_{\rho}S(q;\mu)\frac{\Gamma_{\sigma}(q,p)}{2}~,
    \label{eq:2a.2}
\end{eqnarray}
where $k\equiv p-q$, and $\Lambda_{\rm UV}$ is the ultraviolet cut-off which we take infinity. $D_{\rho\sigma}(p-q)$ is the gluon propagator independent of Landau gauge as
\begin{eqnarray}
D_{\rho\sigma}(p-q) = k^{-2}\left(\delta_{\rho\sigma}-\frac{k_{\rho}k_{\sigma}}{k^2}\right)~,
\label{eq:2a.3}
\end{eqnarray}
$\Gamma_{\sigma}(q,p)$ is a quark-gluon vertex at zero chemical potential. We take a widely used rainbow approximation as
\begin{eqnarray}
    \Gamma_{\sigma}(q,p) = \gamma_{\sigma}~,\label{eq:2a.4}
\end{eqnarray}
and ${\mathcal G}$ is a effective
quark-quark interaction, which we choose a Gaussian-type effective interaction as~\citep{2002PhRvD..65i4026A}
\begin{eqnarray}
{\mathcal G}\left(k^2;\mu\right) = \frac{4\pi D^2}{\omega^6}k^2\exp\left[-\frac{\left(k^2+\alpha\mu^2\right)}{\omega^2}\right]~, \label{eq:2a.5}
\end{eqnarray}
with $\omega=0.5~{\rm GeV}$, and $D = 1~{\rm GeV}^2$. $\alpha$ is a parameter that controls the degree of asymptotic freedom, and we choose a typical value $\alpha=1$~\citep{2011PhRvD..84j5023C}. Note that this approximation reproduces a non-interacting quark matter at finite $\mu$ for $\alpha=\infty$.

Although the rainbow approximation enables us to study the self-energy Eq.~(\ref{eq:2a.2}) regardless of the remaining terms of Eq.~(\ref{eq:2a.1})~\footnote{This merit arises due to the quenched approximation, which neglects fermion loop contributions to the vacuum polarisation as clearly seen in Eq.~(\ref{eq:2a.3}).}, gauge covariance is lost, which leads to an unphysical solution of $S(p;\mu)$. To avoid this, we fix the renormalization constants $Z_1=Z_2=1$, using Ward identity. 

Once the quark propagator $S(p;\mu)$ is obtained, the number density, energy per baryon, and pressure of each quark ($q=u,d,s$) can be obtained as
\begin{eqnarray}
f_q(|\bm{p};\mu|)&=&\frac{1}{4\pi}\int_{-\infty}^{\infty}dp_4{\rm tr}_D\left[-\gamma_4S(p;\mu)\right]\label{eq:2a.6}\\
    n_q(\mu) &=& 2\cdot3\int\frac{d^3p}{\left(2\pi\right)^3}f_q(|\bm{p};\mu|)\label{eq:2a.7}\\
    \epsilon_q(\mu_q)&=&2\cdot3\int\frac{d^3p}{\left(2\pi\right)^3}\sqrt{m_q^2+p^2}f_q(|\bm{p};\mu|)\label{eq:2a.80}\\
P_q(\mu_q)&=&P_q(\mu_{q,0})+\sum_{q=u,d,s}\int_{\mu_{q,0}}^{\mu_q}d\mu n_q (\mu)~\label{eq:2a.8},
\end{eqnarray}
where the traces are applied only to spinor indices. $\mu_{q,0}$ is the theoretically arbitrary value, which we take as $\mu_{u,0}=\mu_{d,0}=0$ and $\mu_{s,0}=460~{\rm MeV}$ following Ref.~\cite{2011PhRvD..84j5023C}. $P_q(\mu_{q,0})$ is the pressure of quark matter in the vacuum, which can be regarded as the effective bag constant
\begin{eqnarray}
B_{\rm DS} = -\left[P_u(\mu_{u  ,0})+P_d(\mu_{d,0})+P_s(\mu_{s,0})\right]
\label{eq:2a.9},
\end{eqnarray}
Following Ref.~\cite{2011PhRvD..84j5023C}, we take $B_{\rm DS}=90~{\rm MeV~fm^{-3}}$.
\begin{figure*}
    \centering
    \begin{minipage}{0.54\linewidth}
\includegraphics[width=\linewidth]{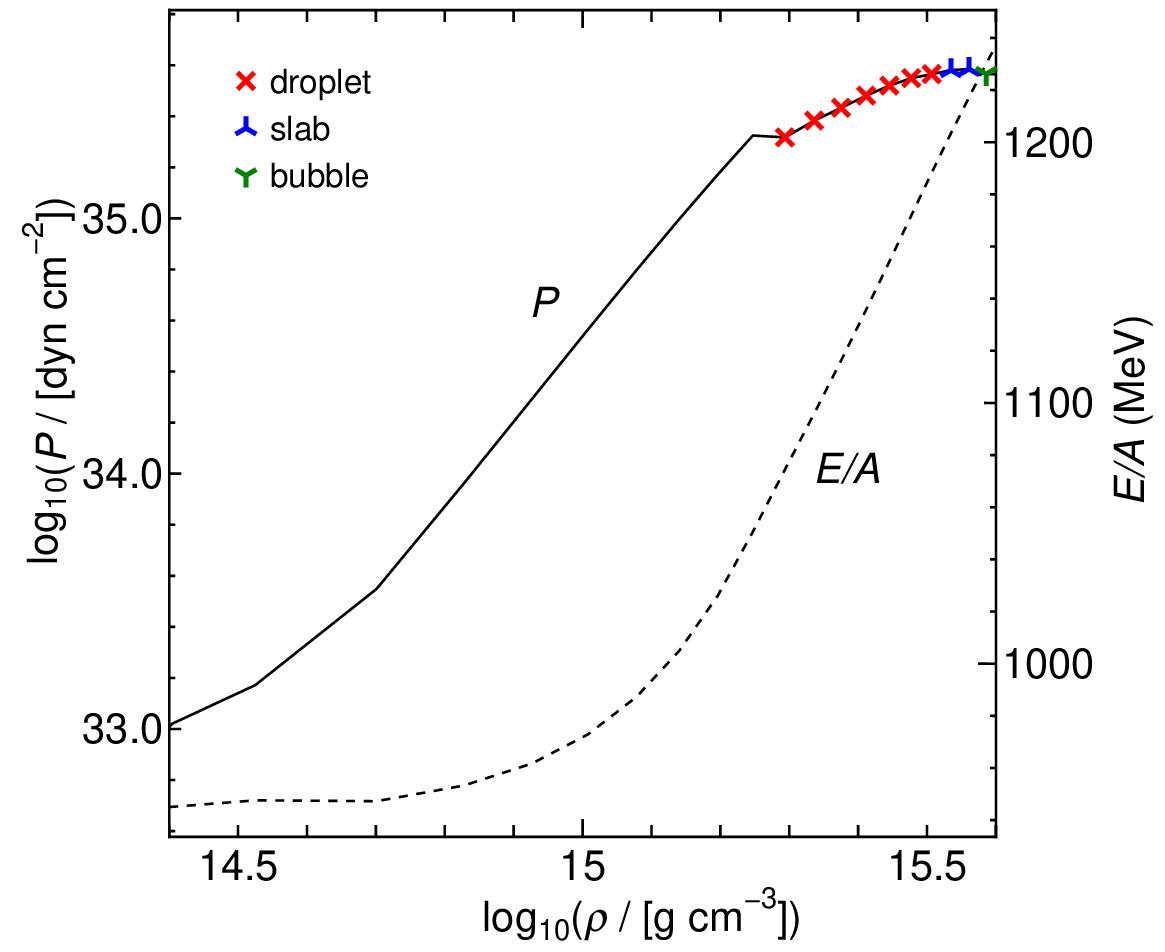}
\end{minipage}    
\begin{minipage}{0.45\linewidth}
\includegraphics[width=\linewidth]{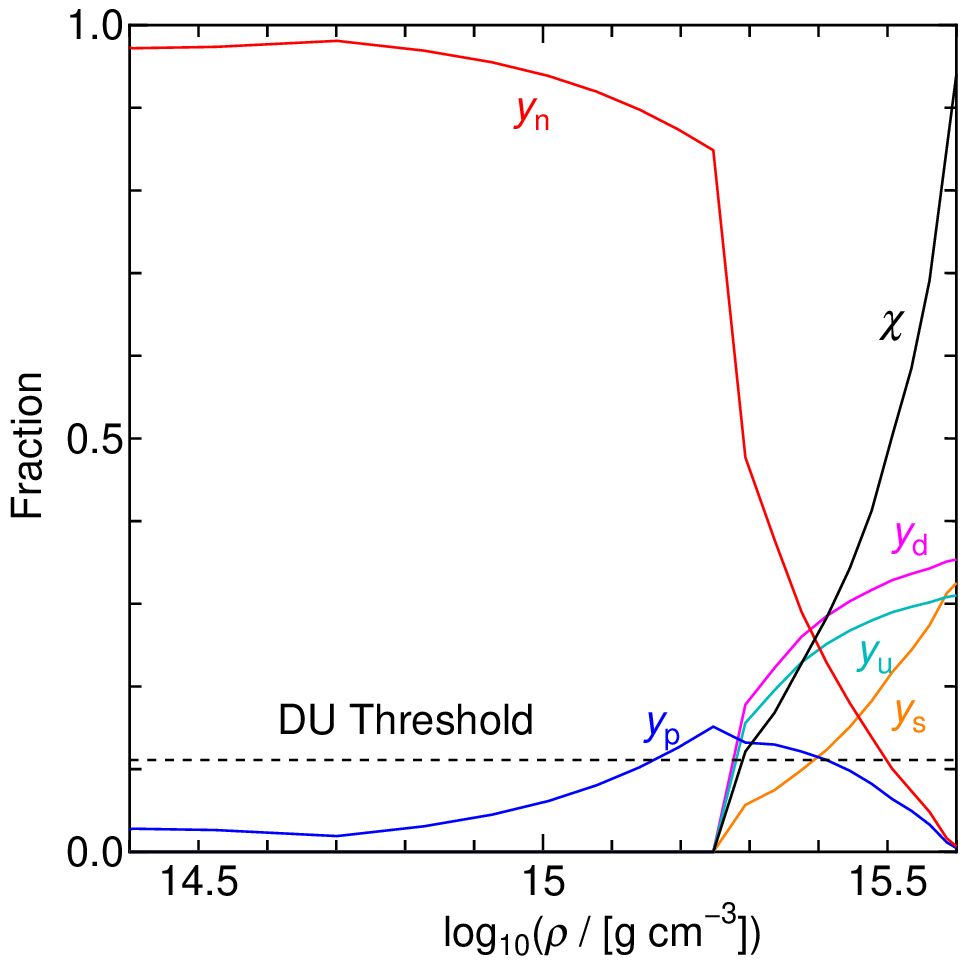}
\end{minipage}    
    \caption{Left: Energy-density dependence of pressure and energy per baryon. The inhomogeneous mixed phases are also shown as different symbols. Right: Energy-density dependence of each particle fraction and quark volume fraction $\chi$. The critical proton fraction for the appearance of the DU process is also plotted.}
    \label{fig:eos}
\end{figure*}

To obtain the EOS of NS matter, we should impose the conditions of neutrino-less beta equilibrium and charge neutrality. In hadronic-matter EOS, which contains two baryons $(n,p)$ and electrons\footnote{Muons are not contained in our EOS, which does not change the EOS and neutrino emissivity significantly at least in cold NSs ($T\lesssim0.1~{\rm MeV}$); The maximum mass is reduced by $\sim \mathcal{O}(0.01)~M_{\odot}$ by including muons. Even if the muon-induced DU process ($p+\mu\rightarrow n+\nu_{\mu},~n\rightarrow p+\mu+\nu_{\mu}$) occurs, the electron-induced DU process has already occurred because of being more electrons than muons, and that it is stronger than muon-induced one.}, these are given as

\begin{eqnarray}
    \mu_n - \mu_p &=&\mu_e~\label{eq:2a.10},\\
Y_p &=& Y_e~\label{eq:2a.12}
\end{eqnarray}
where $\mu_i$ is the chemical potential. On quark-matter EOS, which contains three-flavour $(u,d,s)$ and electrons, these are given as
\begin{eqnarray}
    \mu_d &=& \mu_u + \mu_e~\label{eq:2a.13},\\
    \mu_d &=& \mu_s~\label{eq:2a.14},\\
\frac{2Y_u-Y_d-Y_s}{3} &=& Y_e~\label{eq:2a.16}.
\end{eqnarray}
To construct the hybrid star EOS, we assume the first-order phase transition between the hadron and quark phases. Then, we apply the Gibbs construction, which allows the co-existence of hadron and quark phases with the pressure $P^M$. Such a mixed phase is realized when the following thermodynamical conditions are satisfied:
\begin{eqnarray}
    P^M\equiv P^H&=&P^Q~\label{eq:2a.17},\\
    \mu^H_e &=& \mu^Q_e~\label{eq:2a.18},
\end{eqnarray}
where $H$ and $Q$ denote hadronic and quark phases, respectively. In addition to Eqs. (\ref{eq:2a.10}) and (\ref{eq:2a.13}), and (\ref{eq:2a.14}), the following conditions of beta equilibrium, i.e., the equality of baryonic chemical potential between hadron and quark phases, should be satisfied in mixed-phase:
\begin{eqnarray}
    \mu_n&=&\mu_u + 2\mu_d~~\label{eq:2a.19},\\
         \mu_p &=& 2\mu_u+\mu_d~~\label{eq:2a.20},
\end{eqnarray}
Against the only six equations to describe beta equilibrium, there are seven variables, i.e., $Y_n,Y_p,Y_e,Y_u,Y_d, Y_s$, and $\chi$. A last equation to close the system derives from the global charge neutrality as
\begin{eqnarray}
        \left(1-\chi\right)Y_p + \frac{\chi}{3}\left(2Y_u-Y_d-Y_s\right) &=& Y_e \label{eq:2a.21}
\end{eqnarray}
where $\chi\in\left[0,1\right]$ is the volume fraction of quark matter.

To include the finite-size effects such as the Coulomb interaction and the
surface tension, we employ the Wigner-Seitz approximation in the same manner as in Ref.~\cite{2009PhRvD..80l3009Y} with the surface tension parameter $\sigma=40~{\rm MeV~fm^{-2}}$ (see also our review \cite{2012arXiv1208.0427Y}). Namely, the whole space is sharply divided into the mixed-phase equivalent cells with given geometrical symmetry, which is characterized by the dimensionality $d$; $d=1,2,3$ corresponds to the droplet (or bubble), rod (or tube), and slab, respectively. Thus, the volume fraction is defined as
\begin{eqnarray}
    \chi=\left(\frac{r_d}{r_{WS}}\right)^d~,
\end{eqnarray}
 where $r_{WS}$ and $r_d$ denote the size of the Wigner-Seitz cell and each geometric structure. These properties are reflected in the total internal energy, which is given in the Thomas-Fermi approximation~\footnote{Due to the finite-size effects, the change neutrality, baryon density, and energy density, i.e., Eqs.~(\ref{eq:2a.21}), (\ref{eq:2a.22}), and (\ref{eq:2a.23}), respectively, are slightly modified (for the explicit formulae, see e.g., Ref. \cite{2023arXiv230813973M}), but we pay special attention to the CSC, not the description of finite-size effects in our paper, although they are fully taken into account for our EOS.}. The volume fraction $\chi$ and the corresponding inhomogeneous mixed phase are shown in the left panel of FIG.~\ref{fig:eos}. As the density increases, we can see the transition from droplet, slab, and bubble phases. We note that our EOS does not realize the uniform quark phase even with the maximum mass.

In the end, we can obtain each particle fraction as a function of baryon density $\rho^M_B$ or energy density $\rho^M$ in mixed-phase, which can be obtained from
\begin{eqnarray}
    \rho^M_B&=&\left(1-\chi\right) \rho^H_B + \chi \rho^Q_B ~,
    \label{eq:2a.22}\\
    \rho^M&=&\left(1-\chi\right) \rho^H + \chi \rho^Q~.
    \label{eq:2a.23}
\end{eqnarray}
Information of our EOS is summarized in FIG.~\ref{fig:eos}, such as the energy density, pressure, and fraction of each particle in $\beta$-equilibrium cold matter. Our models show the first-order phase transition beyond $\rho_t\simeq 1.77\times10^{15}~{\rm g~cm^{-3}}$. In our EOS, the DU process is allowed to occur beyond $\rho_{\rm DU}=1.46\times10^{15}~{\rm g~cm^{-3}}$.

\begin{figure}[t]
    \centering
\includegraphics[width=0.6\linewidth]{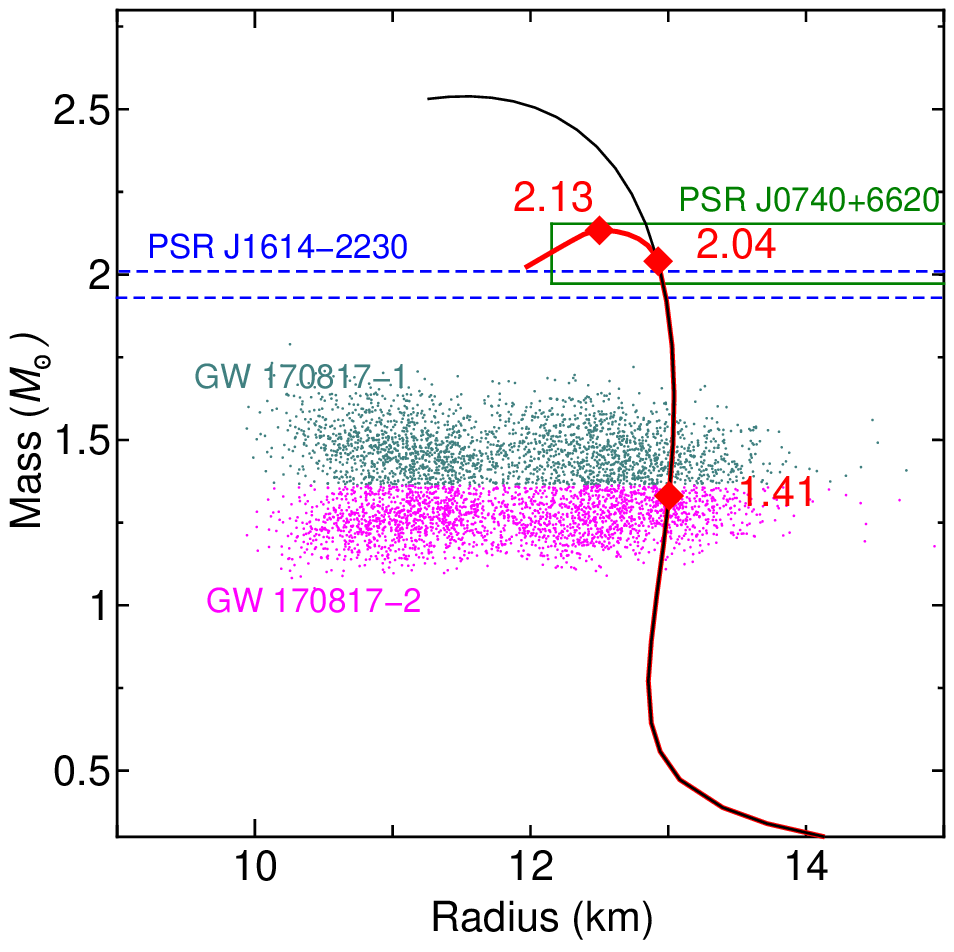}
\caption{Mass-radius relation in our hadronic EOS (black) and hybrid EOS (green). The chosen masses for our cooling calculation are marked as a diamond symbol. We also plot the constraints from the measured tidal deformability of two NSs before they coalesce each other (GW 170817)~\citep{2018PhRvL.121p1101A}, and observed two massive pulsars J1614$-$2230~\citep{2010Natur.467.1081D} and J0740$+$6620~\cite{2021ApJ...918L..28M}.}
    \label{fig:mr}
\end{figure}

Once the EOS is constructed, one can obtain the mass-radius relation through the TOV equation~\citep{1939PhRv...55..364T,1939PhRv...55..374O}, which is shown in FIG.~\ref{fig:mr}. As reported in Ref.~\cite{2016JPhCS.665a2068Y}, our hybrid EOS can account for the observations of massive $2~M_{\odot}$. The radius around the canonical mass $1.4~M_{\odot}$ is 13 km, which is consistent with many observational constraints such as GW 170817 and NICER observation of PSR J0740$+$6620.

\subsection{Neutrino Emission Processes and Superfluidity / Color Superconductivity}

In the hadronic phase, we consider the \textit{conventional} cooling processes (see Ref.~\cite{2001PhR...354....1Y} for their list), which are the same as our previous work~\citep{2019PTEP.2019k3E01D,2022IJMPE..3150006D}: Slow neutrino cooling processes mainly composed of modified Urca, and bremsstrahlung, are always open to occur in any NSs. These emissivities are approximately 
\begin{eqnarray}
\epsilon_{\nu}^{\rm Slow} \approx 10^{19-21}T_9^8~{\rm erg~cm^{-3}~s^{-1}}~,\label{eq:nu1}
\end{eqnarray}
where $T_9$ is the local temperature in a unit of $10^9$ K. 

In addition to slow cooling processes, when the temperature becomes lower than the superfluid transition temperature $T_c$, the pair breaking and formation (PBF) process occurs as the release of thermal energy. The emissivity is given as \cite{1995A&A...297..717Y,2004ApJS..153..269K}:
\begin{eqnarray}
\epsilon^{\rm PBF}_{\nu} \approx 10^{21-22}T_9^7\tilde{F}_i(T/T_{\rm cr}) {\rm erg~cm^{-3}~s^{-1}}\,, \label{eq:nu2}
\end{eqnarray}
where $\tilde{F}_i(T/T_{\rm cr})$ indicates the efficiency of PBF processes s a function of $T/T_{\rm cr}$ with ${}^1S_0$ for $i=s$ and ${}^3P_2$ baryons for $i=t$. The concrete formula of the \textit{control function} is given as \cite{1999A&A...343..650Y}:
\begin{eqnarray}
\tilde{F}_i &=& \frac{1}{4\pi}\int d\Omega~y^2\int_0^{\infty}dx \frac{z^4}{1+e^z}~ \label{eq:nu4},
\end{eqnarray}
with $z=\sqrt{x^2+y^2}$, $y=k_i\frac{T_{\rm cr}}{T}$, where $k_i$ is the conversion constant factor with the state $i$, and $\int d\Omega$ denotes the angle averaging integration. $\tilde{F}_i$ has a maximum value (unity) for $T\sim0.65(0.5)$ for ${}^{3}P_2$(${}^{1}S_0$) state, respectively, and zero for $T\lesssim~0.2 T_{\rm cr}$. As we see the coefficients of Eqs. (\ref{eq:nu1}) and (\ref{eq:nu2}), the PBF process is generally stronger than slow cooling processes except for cold NSs.

We also consider the nucleon DU process, i.e., neutrino emissions through $\beta$ decay and inverse-$\beta$ decay:
\begin{eqnarray}
p+e\rightarrow n+\nu_e,~n\rightarrow p+e+\bar{\nu}_e.~\label{eq:nu5}
\end{eqnarray}
The DU process usually occurs in heavy NSs, and the emissivity is given as 
\begin{eqnarray}
\epsilon^{\rm DU}_{\nu} = 4.0\times 10^{27} \left(\frac{m_N^*}{m_N}\right)^2 \left(\frac{\rho}{\rho_{\rm nuc}} \right)^{2/3} T_9^6~{\rm erg~cm^{-3}~s^{-1}}\,, \label{eq:nu6}
\end{eqnarray}
where $m_N^*/m_N$ is the effective mass ratio of nucleons, assuming equal masses of neutrons and protons. Compared with the standard cooling processes, the DU process is strong as we see the coefficient in Eq.~(\ref{eq:nu5}), since all observed NSs are cold enough to satisfy $T_9\lesssim1$ (but maybe except for the NS 1987A~\citep{2020ApJ...898..125P,2023ApJ...949...97D}). However, the DU process is realised under conditions, where the proton fraction $Y_p$ exceeds $1/9$ with no muons present, because the momentum conservation between neutrons, protons, and electrons is satisfied.
In our EOS, this corresponds to $M_{\rm DU}\ge 1.70M_{\odot}$, where $M_{\rm DU}$ is the threshold of the mass where the DU process operates. 

In the quark phase, we assume 2-flavour case for neutrino emissivity \footnote{We assume that the effect of $s$-quarks is not significantly large. The s-quarks decay into u-quarks which is in superconducting state (2SC component of 2SC+$\braket{dd}$ phase), and the emissivity caused by $s$-quarks is strongly suppressed. Also, the effect of Cabibbo suppression is significantly large compared to $ud$-quark $\beta$ decay. To demonstrate the effect of 2SC+$\braket{dd}$ phase, we assume the simplest quark model which consists of $u$ and $d$ quarks.}. The dominant cooling process in unpaired quark matter is the
quark $\beta$ decay of
\begin{eqnarray}
u+e\rightarrow d+\nu_e,&&~d\rightarrow u+e+\bar{\nu}_e.~~\label{eq:nu7}
\end{eqnarray}
The emissivity in the case of massless non-interacting up and down quarks is given by \cite{1980PhRvL..44.1637I,1982AnPhy.141....1I}:
\begin{eqnarray}
\epsilon_{\nu}^{\rm q\beta d} \simeq 8.8\times10^{26}\alpha_s\left(\rho/\rho_{\rm nuc}\right)Y_e^{1/3} T_9^6~{\rm erg~cm^{-3}~s^{-1}},~~\label{eq:nu8}
\end{eqnarray}
where $\alpha_s$ is the strong coupling constant and is fixed to be 0.1.
This emissivity is similar to that of the DU process, and enough to cool NSs rapidly. 

If quarks are paired, since most of them cannot contribute to the quark $\beta$ decay, the emissivity is reduced as a function of $\exp(-T_{{\rm cr},q}/T)$ (e.g., ~\cite{2001A&A...368..561B}), where $T_{{\rm cr},q}$ is the critical temperature of CSC and assumed to be 10 MeV. In the 2SC phase, the emissivity is almost the same as Eq. (\ref{eq:nu8}) multiplied by 1/3. In the CFL phase, on the other hand, the quark $\beta$ decay is highly suppressed and practically invalid. This can be applied to quark-modified Urca and quark bremsstrahlung processes, which are negligible in all cases because we consider cold NSs ($T_9\lesssim1$) and the suppression factor is quite small as $\exp(-2T_{{\rm cr},q}/T)$. Instead, the dominant cooling process in quark CFL matter is electron-electron scattering as\footnote{In the CFL, the scattering of gluons appearing as the Numbu-Goldstone boson could be a cooling process due to the production of neutrinos, but it is quite sensitive to temperature~\cite{2002PhRvD..66f3003J,2003NuPhA.714..337R}, and little affects cooling curves with $T\lesssim10~{\rm MeV}$.}
\begin{eqnarray}
    e+e\rightarrow e+e+\nu+\bar{\nu},~\label{eq:eqnu9}
\end{eqnarray}
whose neutrino emissivity is given as \cite{1999AcPPB..30.1125K}:
\begin{eqnarray}
\epsilon_{\nu}^{ee}=2.8\times10^{12}\left(\rho/\rho_{\rm nuc}\right)Y_e^{1/3} T_9^8~{\rm erg~cm^{-3}~s^{-1}}.~~\label{eq:nu10}
\end{eqnarray}
In our EOS, baryons and electrons exist even in the maximum central density regions. Since the quark $\beta$ decay is practically invalid due to CFL pairing, baryonic cooling processes are dominant for total neutrino emissivity.

The total neutrino emissivity in the quark-hadron mixed phase is the summation of individual emissivities shown above. Some processes have conditions and some are affected by the superfluidity or the superconductivity. Considering the quark CSC and nucleon superfluidity, and dropping significant small terms, the total neutrino emissivity can be written as

\begin{align}
    \epsilon_\nu^\mathrm{Total} &\simeq (1-\chi)\epsilon_\nu^\mathrm{PBF}\tilde{F}_t\nonumber\\ &+ f_\mathrm{sf} \biggl( \left(1-\chi\right)\epsilon_\nu^\mathrm{DU}\Theta\left(\rho-\rho_{\rm DU}\right)\nonumber \\ &+ \chi \frac{1}{3} \epsilon_\nu^\mathrm{q\beta d}\Theta\left(\rho-\rho_t\right)\biggr),
\end{align}

where $f_\mathrm{sf}$ is the suppression factor by superfluidity. In very low-temperature regions compared to $T_{\rm cr}$, it is given as~\cite{1998PhR...292....1T} 
\begin{align}
    f_\mathrm{sf} &= \exp\left(\frac{-8.4 T_\mathrm{cr}}{T}\right)~.
\end{align}
Here, we drop some negligible terms of neutrino emissivities of slow cooling, electron scattering, and quark $\beta$ decay of 2SC-paired quarks, the last of which becomes much weaker due to the large CSC critical temperature. $\tilde{F}_t(T/T_{\rm cr})$ indicates the efficiency of PBF processes with ${}^3P_2$ neutrons as a function of $T/T_{\rm cr}$, as given by Eq.~(\ref{eq:nu4}). $\Theta$ is the Heaviside function in respect to the density. On the other hand, low-mass NSs with $M_{\rm NS}<M_{\rm DU}$ obey the standard cooling scenario mainly composed of Eqs.~(\ref{eq:nu1}) and (\ref{eq:nu2}), which actually depend on the neutron ${}^{3}P_2$ SF models as well. Moreover, the strength of the PBF process also varies with neutron ${}^{3}P_2$ SF models.

\begin{figure}[t]
\centering
\includegraphics[width=0.6\linewidth,clip]{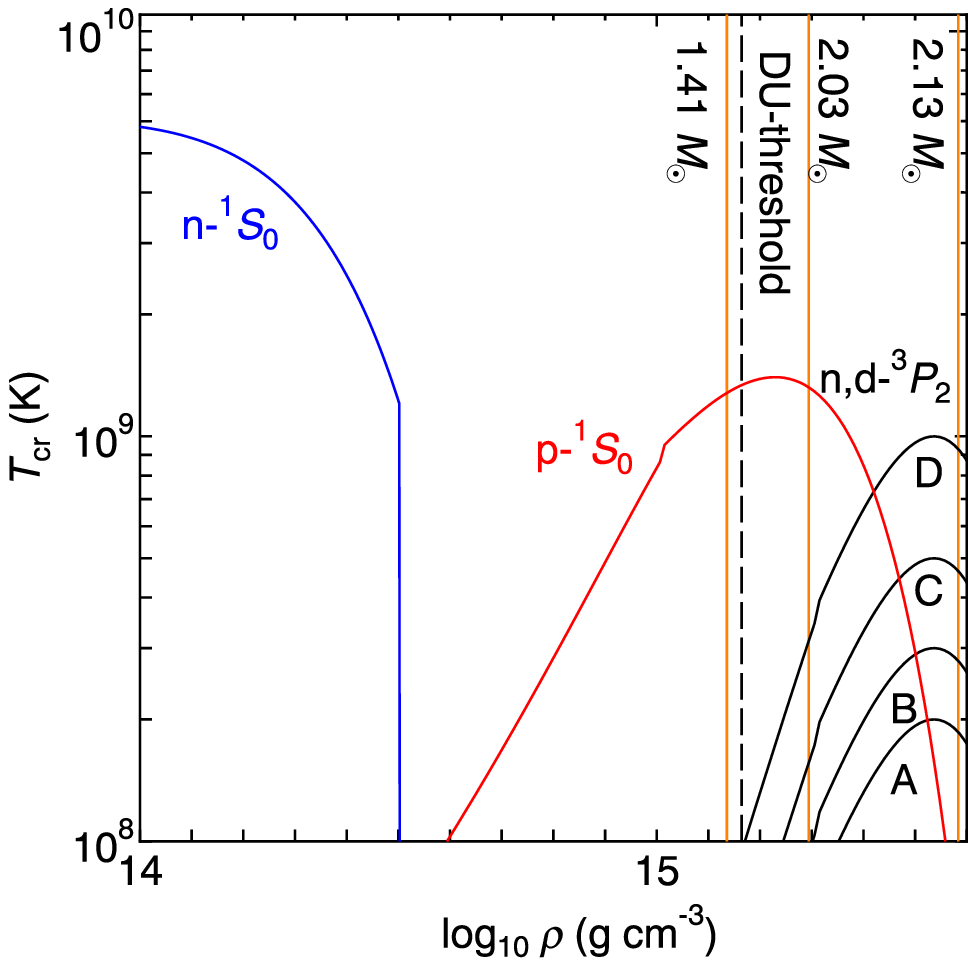}
\caption{Critical temperatures of each superfluid/superconductive state. Black curves show the different ${}^3P_2$ models of neutrons and unpaired $d$ quarks. Note that the CSC state does not appear in this figure due to the high critical temperature as 10 MeV (=$1.16\times10^{11}~{\rm K}$). Central densities corresponding to $M_{\rm NS}=1.41,2.03$ and 2.13 $M_{\odot}$ are drawn as orange lines. The threshold density of the DU process is also drawn as a black line.}
\label{fig:sf}
\end{figure}

FIG.~\ref{fig:sf} displays our superfluid models. We consider the four different SF models with the ${}^3P_2$ state, which controls the strength of 2SC+$\braket{dd}$ as well as the neutrons superfluidity, while other states are fixed. 
Following Fujimoto et al. \cite{2020PhRvD.101i4009F} (see discussion around Eq. (19) in their paper), we set the equal critical temperature of $d$-quark ${}^3P_2$ to that of neutron ${}^3P_2$. Even the neutrons, where the uncertainty of superfluidity is significant (e.g., \cite{2015PhRvC..91a5806H}), the situation is thought to be even more complex when it comes to quarks. Although model A--D is independent of nuclear theories, the maximum critical temperature of model A--D in the ${}^3P_2$ state, $\left(2-10\right)\times10^8~{\rm K}$ is reasonable according to nuclear models~(see FIG. 1 in \cite{2015PhRvC..91a5806H}). In particular, model C is similar to that inferred from
the observed cooling rate of Cassiopeia A~\cite{2011PhRvL.106h1101P}. The critical temperature models of ${}^3P_2$ superfluid of neutrons and $d$-quarks have onset density around $2-3 \times 10^{14} ~\mathrm{g/cm^3}$ close to the model of SYHHP, which is a phenomenological model for Cas A observations \cite{2015PhRvC..91a5806H,2011MNRAS.412L.108S}. They appear at a higher density than theoretical models of neutron ${}^3P_2$, which is, however, highly uncertain in high-density regions.

Note that the impact of ${}^1S_0$ superfluid models on cooling curves is smaller than that of ${}^3P_2$~\footnote{It is possible that, as the density increases, the superconductive phase of protons in the ${}^1S_0$ state may move to the two $ud$ plus $uu$ pairs, 2SC+$\braket{uu}$ phase. However, the impact of the 2SC+$\braket{uu}$ phase seems very minor compared to the 2SC+$\braket{dd}$ because of following reasons:
\begin{itemize}
\item The critical temperature of ${}^1S_0$ is higher typically by 1-2 orders of magnitude than ${}^3P_2$.
\item Neutrons are always more than protons as shown in FIG.~\ref{fig:eos}. 
\end{itemize}
}.

\subsection{Other Setup for NS cooling}
\label{subsec:cool}

We utilize the same neutron-star cooling code as in our previous study~\citep{2013ApJ...765....1N}, which simultaneously solves TOV and thermal transport equations. The latter equation is given on the mass coordinates $M_r$:
($c=G=1$)~\cite{1977ApJ...212..825T,2018A&A...609A..74P}:
\begin{eqnarray}
\frac{\partial (L_{r}e^{2\phi})}{\partial M_{r}} & = &
      -e^{2\phi}\left(\varepsilon_{\nu} + e^{-\phi}C_V\frac{\partial T}{\partial t}
      \right)~, \label{eq:eq5} \\
      \frac{\partial \ln T}{\partial \ln P} & = & \frac{3}{16\pi} \frac{\kappa L_r P}{M_{{\rm t}r}aT^4}\frac{\rho_0}{\rho}\left( 1 + \frac{P}{\rho} \right)^{-1}\times \nonumber \\
&&\left( 1 + \frac{4 \pi r^3 P}{M_{{\rm t}r}} \right)^{-1}\left(1-\frac{2M_{{\rm t}r}}{r}\right)^{1/2} \nonumber \\
&& + \left[1 - \left( 1 + \frac{P}{\rho} \right)^{-1} \right]~, \label{eq:eq6}
\end{eqnarray}
where $M_{tr}$ and $M_r$ are gravitational and rest masses enclosed in a radius $r$; $\rho$ and $\rho_0$ denote the total mass-energy and rest mass densities; $P$, $T$, and $L_r$ are the pressure, local temperature, and local photon luminosity, respectively, $\varepsilon_\nu$ denotes the energy loss rate by neutrino emission; $\phi$ is the gravitational potential in unit mass; $a$ is the Stefan-Boltzmann constant; $C_V$ is the specific heat; $\kappa$ is the opacity. As the boundary condition, we impose  \textit{radiative zero boundary condition} at a sufficiently closed area to the photosphere~\cite{1984ApJ...278..813F}. By solving Eqs.~(\ref{eq:eq5}), (\ref{eq:eq6}) and the TOV equations numerically, we can obtain the time evolution of the luminosity, and the surface temperature via Stefan--Boltzmann law. 

Regarding the surface compositions, we mostly consider the ${}^{56}{\rm Fe}$ surface, corresponding to the absence of light elements onto the NS. However, surface compositions are one of the important ingredients in cooling curves, although they have nothing to do with the properties of interior NSs such as the CSC. For Model C, therefore, we also consider the pure ${}^{4}{\rm He}$ surface, which roughly gives the maximum amount of light elements. 

In terms of age and observational data, we mostly take the same data as in our previous work~\citep{2013ApJ...765....1N}. In particular, the Vela is cold for its age, and its observational data has small uncertainties compared to others. Furthermore, we also consider the observational data of PSR J0205$+$6449 (in supernova remnant 3C58)
\footnote{The supernova remnant 3C 58 and the CCO PSR J0205$+$6449 have been thought to be associated with the historical supernova SN 1181. However, recent analysis of SN 1181 indicates that another nebula Pa30 exists near the position of the supernova, and that the expansion rate of the nebula matches the 1181 event~\cite{2021ApJ...918L..33R}. We adopt the age of 3C 58 derived from the expansion velocity of the remnant rather than from the historical event.}
, PSR B2334$+$61 (often called Vela-like pulsar), and CXOU J0852$-$4617 (or Vela Jr.), which have been very recently estimated based on the analysis of XMM--Newton and Chandra data ~\citep{2024arXiv240405371M}\footnote{The age of 3C58 and Vela Jr. are quite uncertain. The former derives from the significant difference between the historical record and observed expansion rate of the supernova remnant~\cite{2013A&A...560A..18K}. The latter is due to the lack of reliable historical records~\cite{2015ApJ...798...82A,2023A&A...673A..45C}.}. They identified these three cold NSs for their ages even compared to Vela, which is beyond a minimal cooling scenario\footnote{This implication itself has been well known~(e.g., Ref.~\cite{2004ApJS..155..623P}), though no lower bounds of surface temperature of the three NSs were there so far.}. Hence, these four NSs must be strong constraints to probe the compositions of high-density matter.

\section{Results}
\label{sec:results}

\begin{figure*}[t]
\centering
\begin{minipage}{0.4\linewidth}
\includegraphics[width=\linewidth,clip]{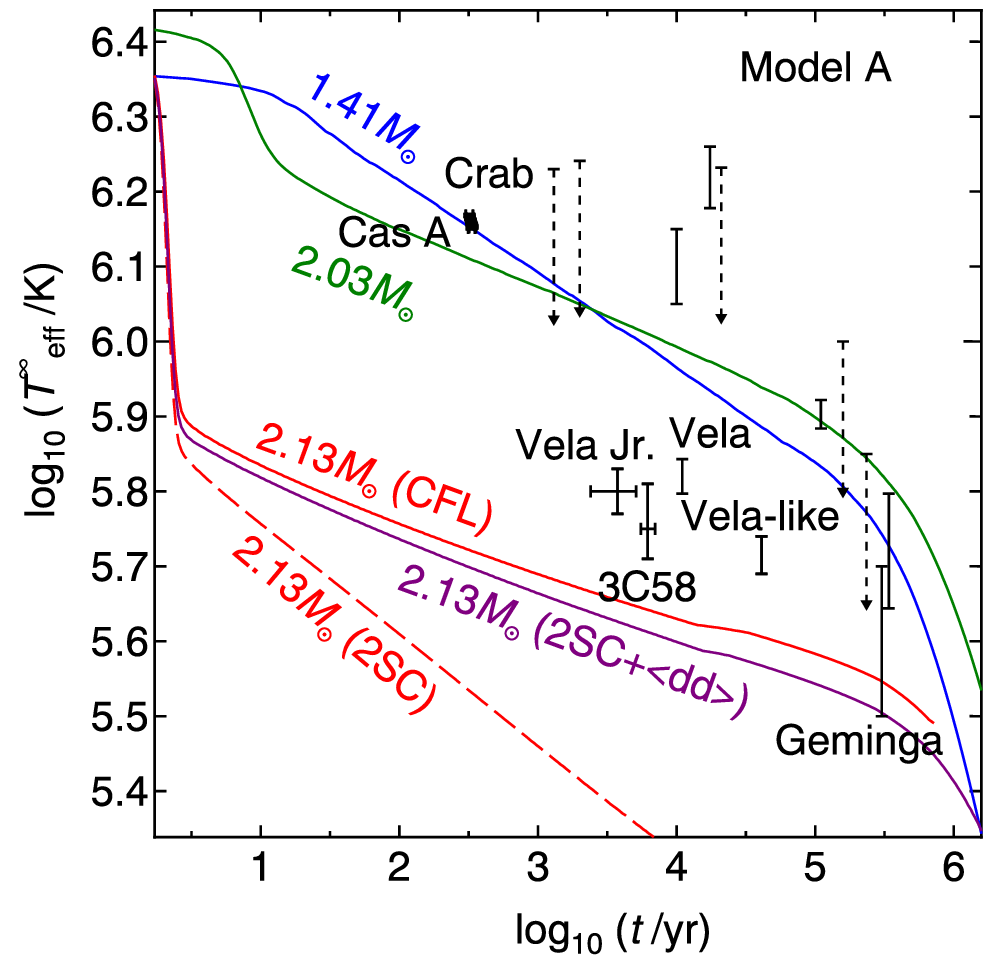}
\end{minipage}
\begin{minipage}{0.1\linewidth}
\end{minipage}
\begin{minipage}{0.4\linewidth}
\includegraphics[width=\linewidth,clip]{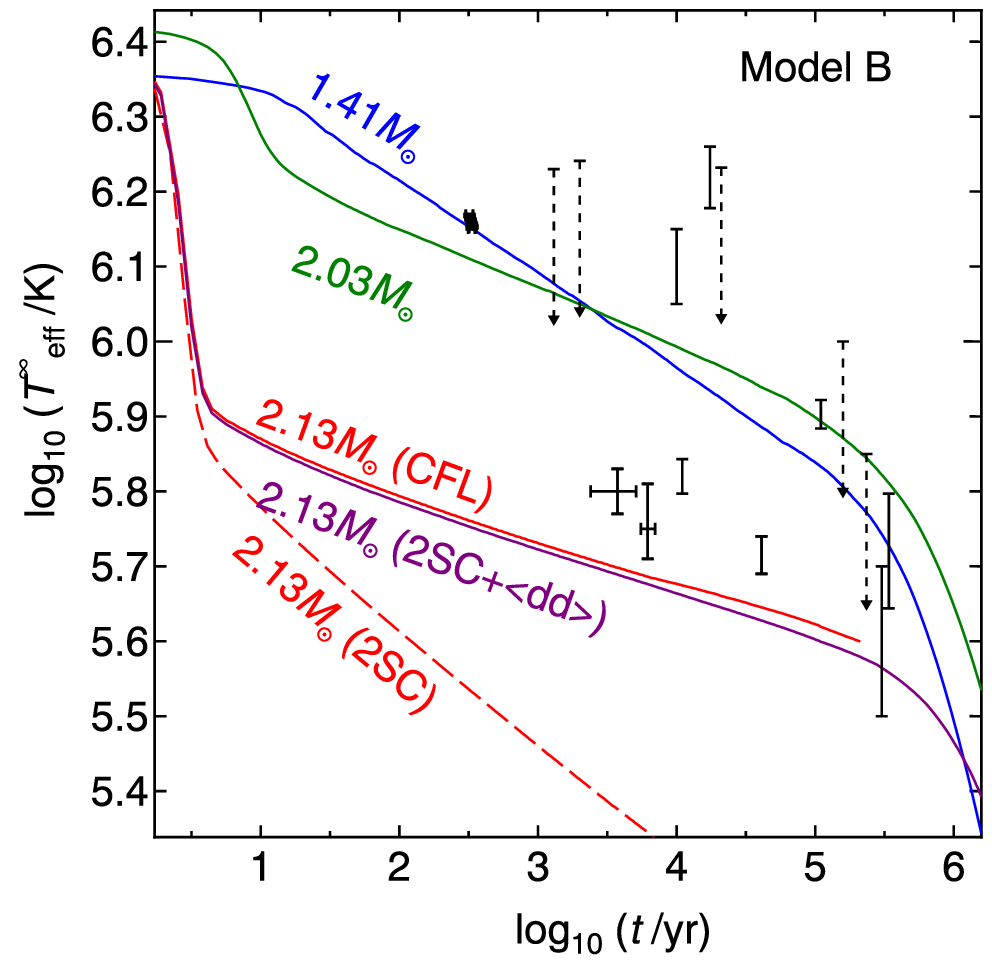}
\end{minipage}

\vspace*{-0cm}

\begin{minipage}{0.4\linewidth}
\includegraphics[width=\linewidth,clip]{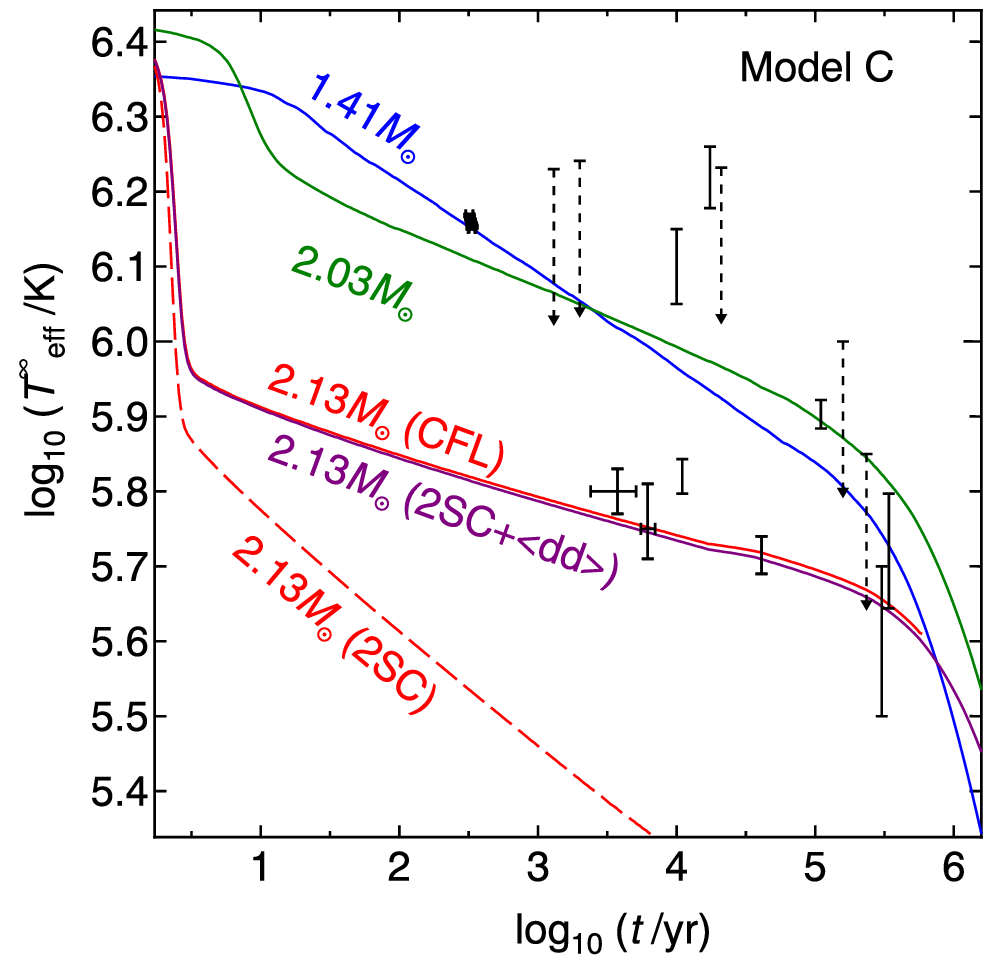}
\end{minipage}
\begin{minipage}{0.1\linewidth}
\end{minipage}
\begin{minipage}{0.4\linewidth}
\includegraphics[width=\linewidth,clip]{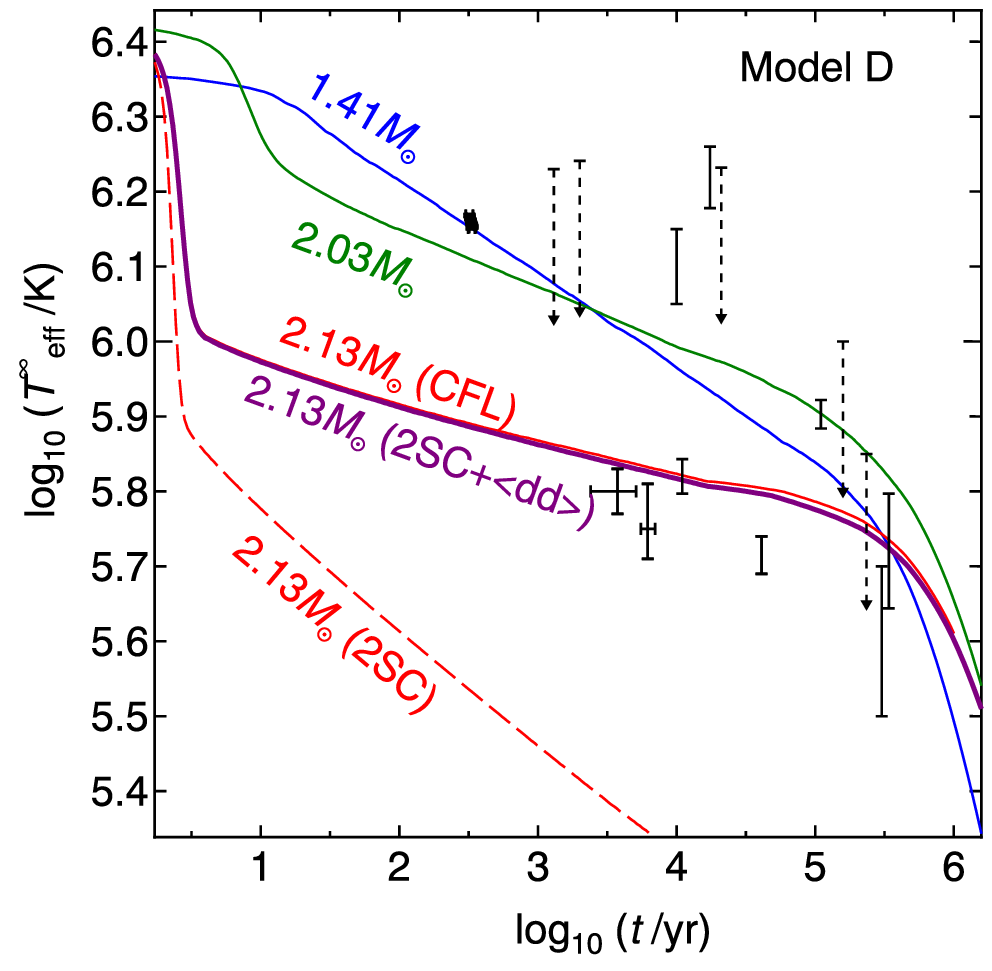}
\end{minipage}

\vspace*{-0cm}
\caption{Cooling curves with various superfluid models. The blue (green) line denotes 1.41$M_\odot$ (2.03$M_\odot$), respectively. The red solid (dashed) line denotes 2.13$M_\odot$ with CSC paring of CFL (2SC), respectively. The purple line denotes 2.13$M_\odot$ with 2SC+$\braket{dd}$ model. Each superfluid model is shown on the right-top corners of each panel.}
\label{fig:cc}
\end{figure*}

\begin{figure}[t]
    \centering
    \includegraphics[width=0.6\linewidth]{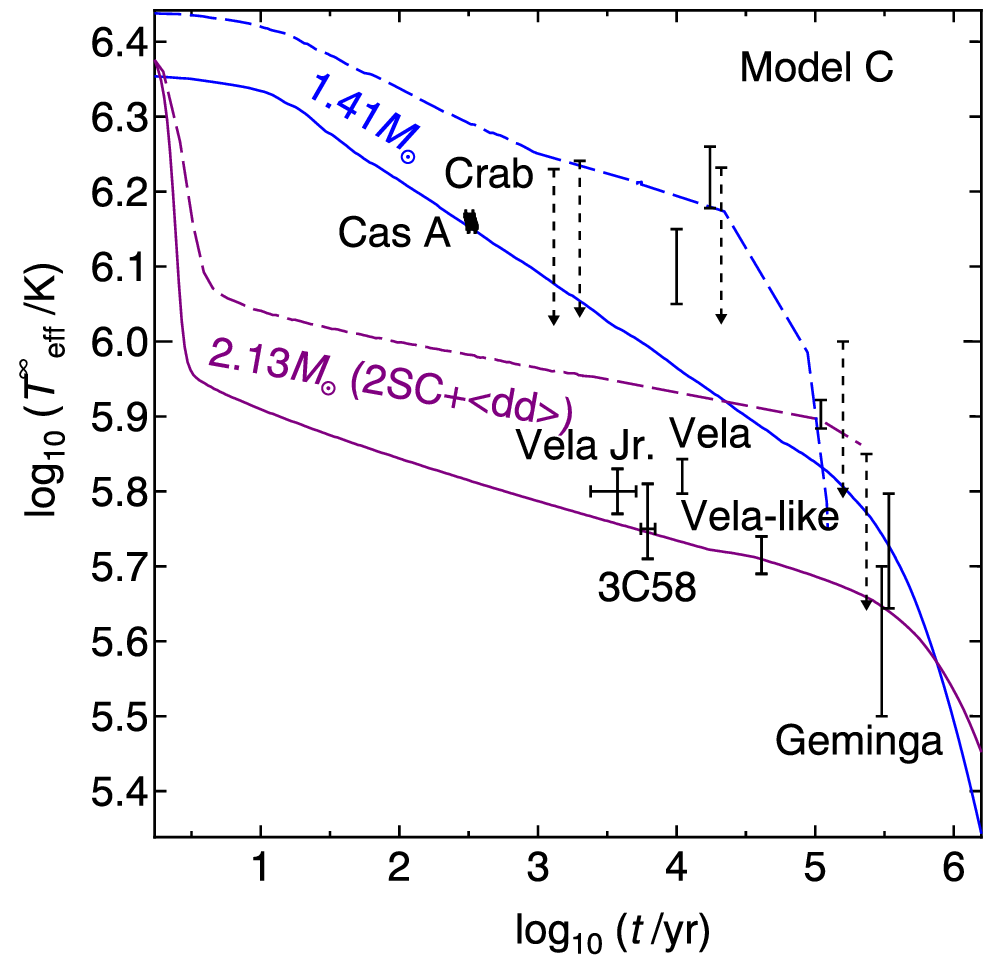}
    \caption{Cooling curves with Model C, $M_{\rm NS}=1.41,~2.13~M_{\odot}$, and pure Fe envelope (solid line) and He envelope (dashed line).}
    \label{fig:cche}
\end{figure}

FIG.~\ref{fig:cc} shows various cooling curves in our hybrid star EOS;
For $2.13~M_{\odot}$ NSs, where the quark phase is allowed to appear, we show three scenarios of CSC of only 2SC phase, only CFL phase, and 2SC+$\braket{dd}$ phase\footnote{For simplicity, we consider the case where either 2SC (or 2SC+$\braket{dd}$) or CFL is allowed, but their combination is possible because they can be smoothly connected in regard with the symmetry~\cite{2020PhRvD.101i4009F}.}.
On the other hand, our models with 1.4 and 2 $M_{\odot}$ NSs are composed of hadronic matter. These light NSs show moderate cooling behavior, as any of the fast cooling processes do not occur. In the early stage ($t<1000~{\rm yr}$), 2 $M_{\odot}$ NSs cools faster than 1.4 $M_{\odot}$, but it becomes opposite in the late stage. This is because the higher-mass models show higher central density, leading to higher neutrino luminosity. And, since the neutrino emissivity is sensitive to the temperature ($\propto T^{7-8}$ in these cases), neutrino luminosity decreases rapidly for higher-mass models. The mass relationship within the standard cooling scenario we show is consistent with many literature~(e.g., Ref.~\cite{2017IJMPE..2650015L}).

\begin{figure*}[t]
    \centering
\begin{minipage}{0.315\linewidth}
\includegraphics[width=\linewidth,clip]{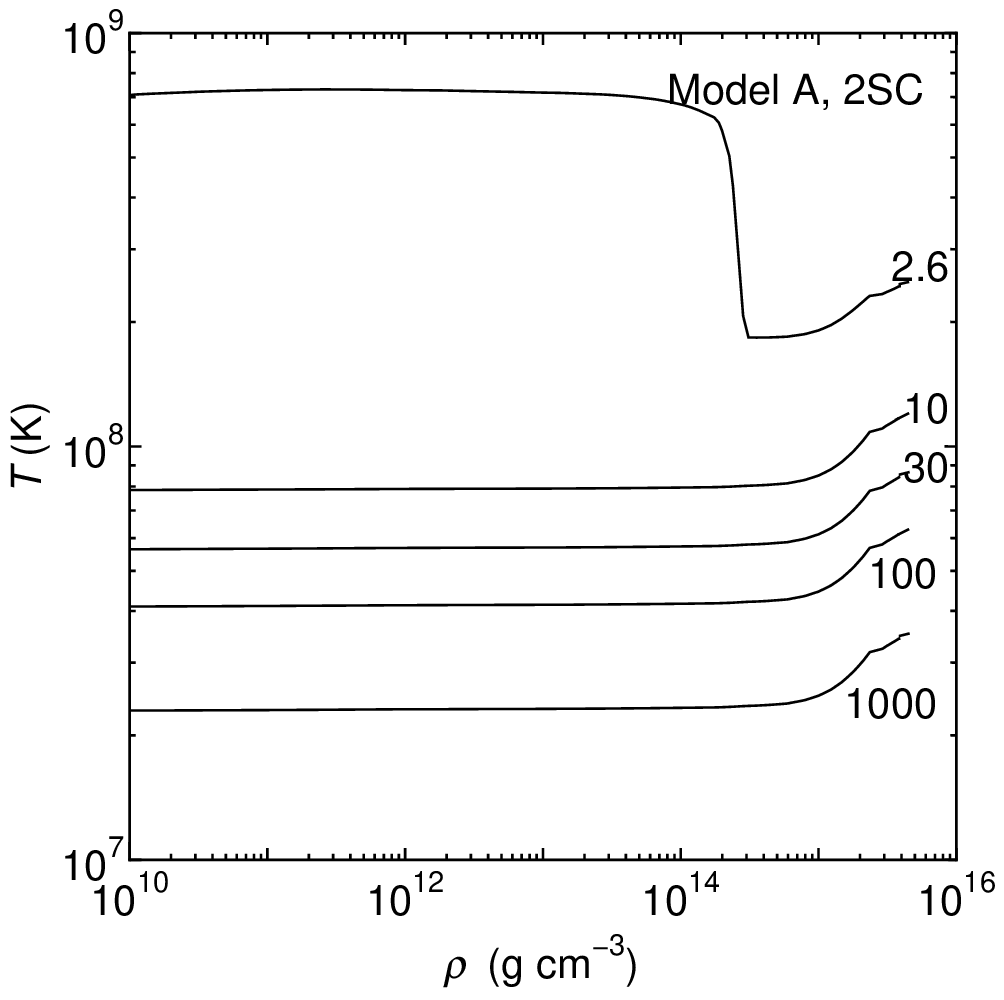}
\end{minipage}
\begin{minipage}{0.315\linewidth}
\includegraphics[width=\linewidth,clip]{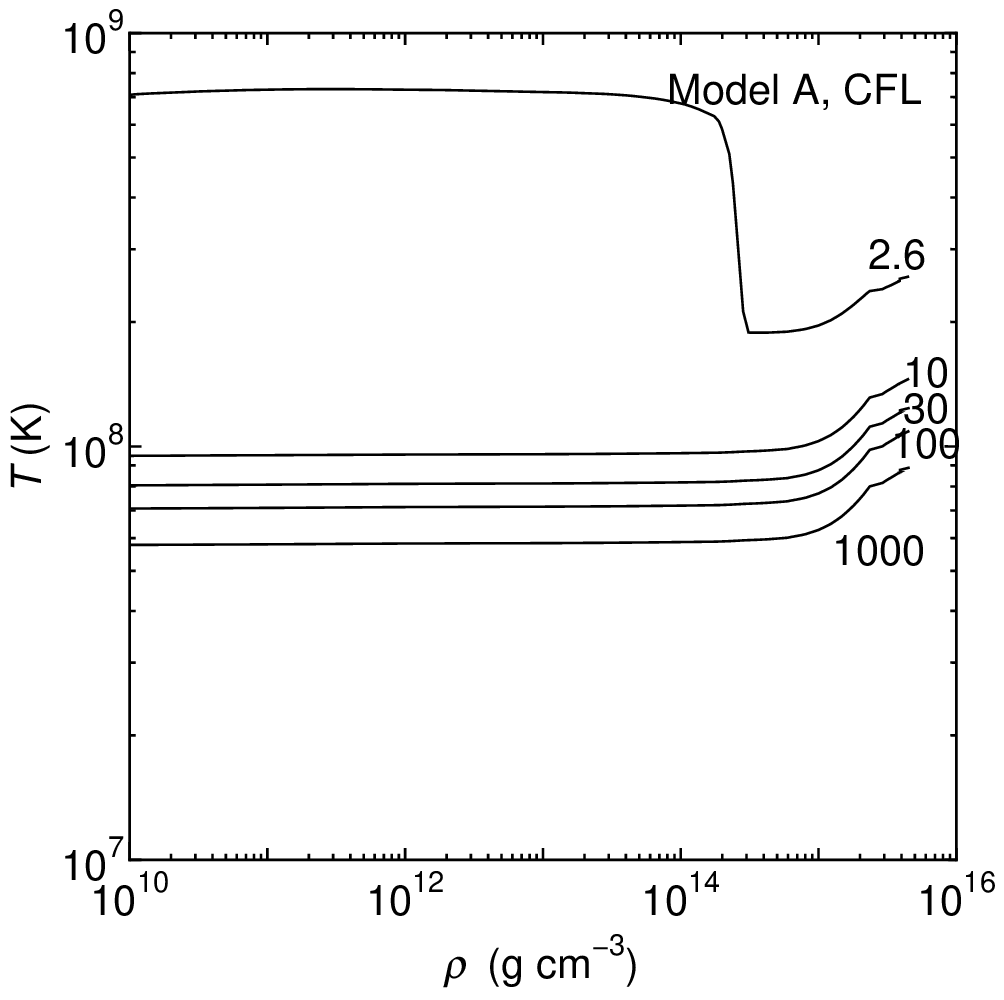}
\end{minipage}
\begin{minipage}{0.315\linewidth}
\includegraphics[width=\linewidth,clip]{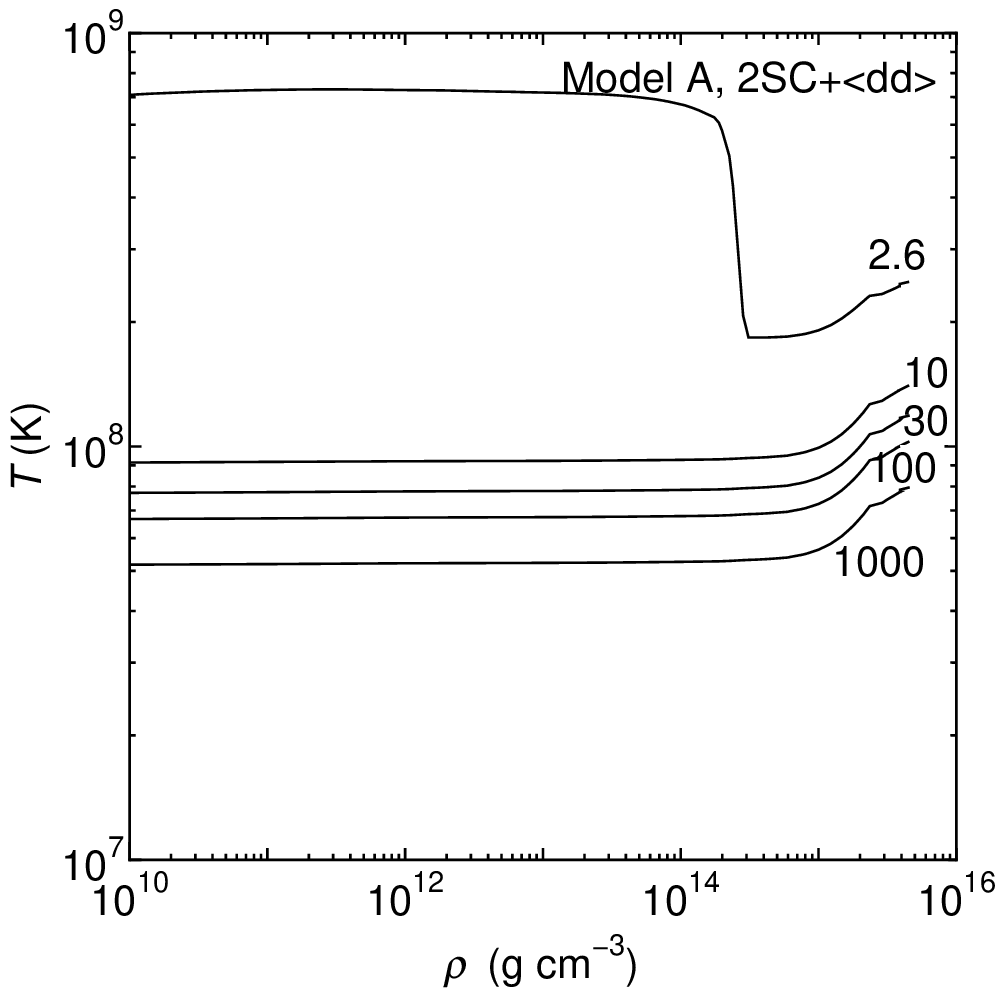}
\end{minipage}
\begin{minipage}{0.315\linewidth}
\includegraphics[width=\linewidth,clip]{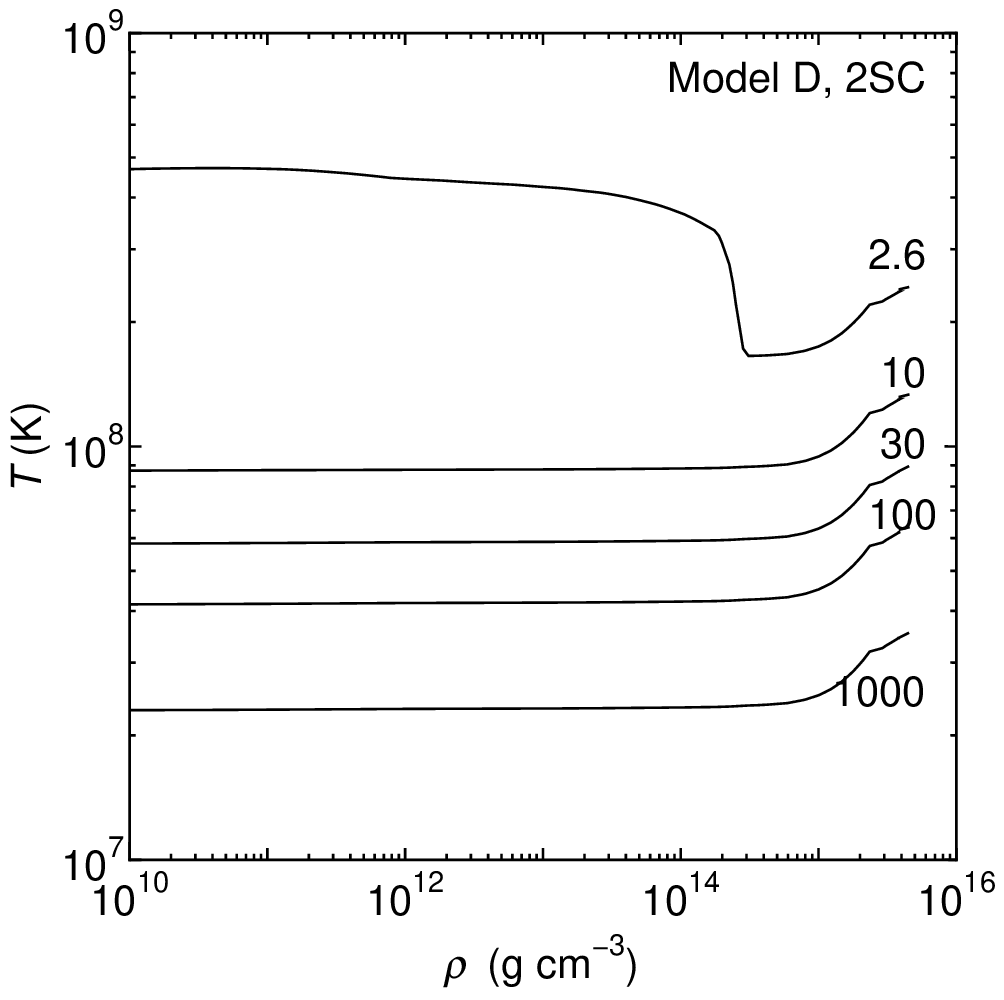}
\end{minipage}
\begin{minipage}{0.315\linewidth}
\includegraphics[width=\linewidth,clip]{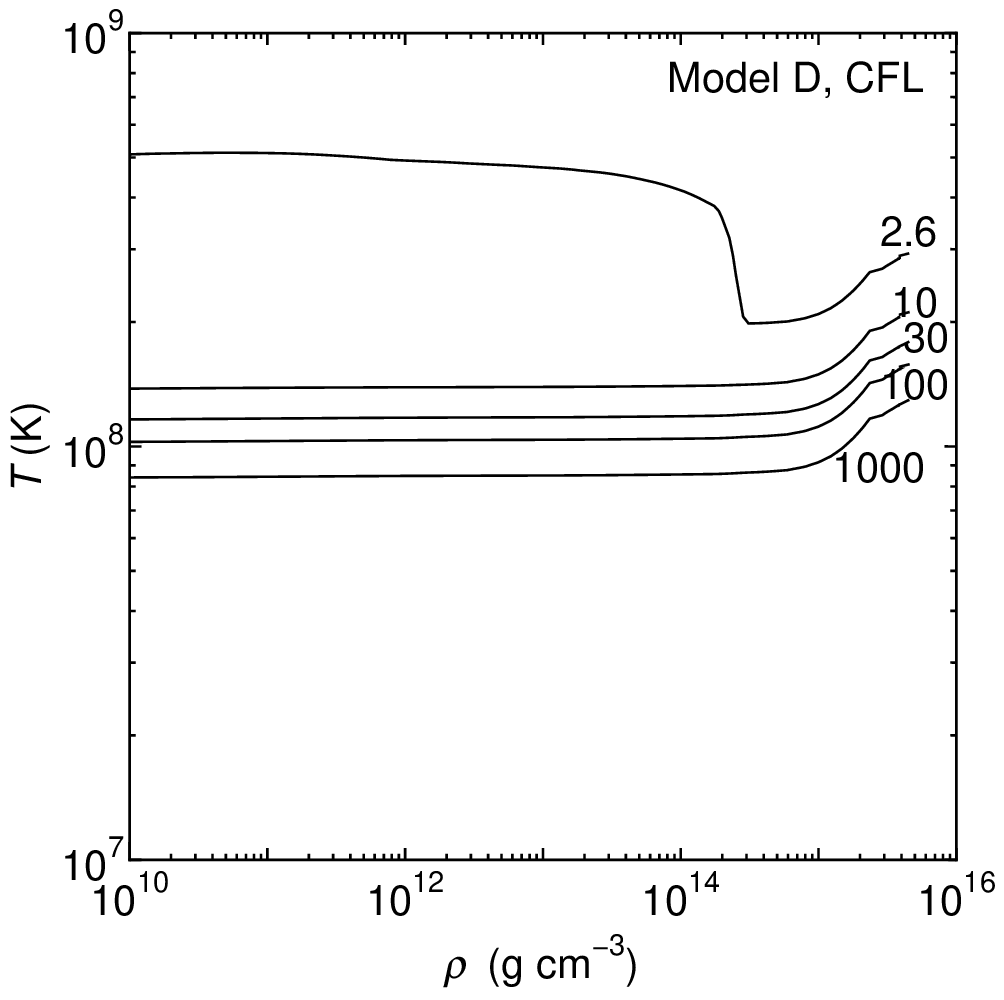}
\end{minipage}
\begin{minipage}{0.315\linewidth}
\includegraphics[width=\linewidth,clip]{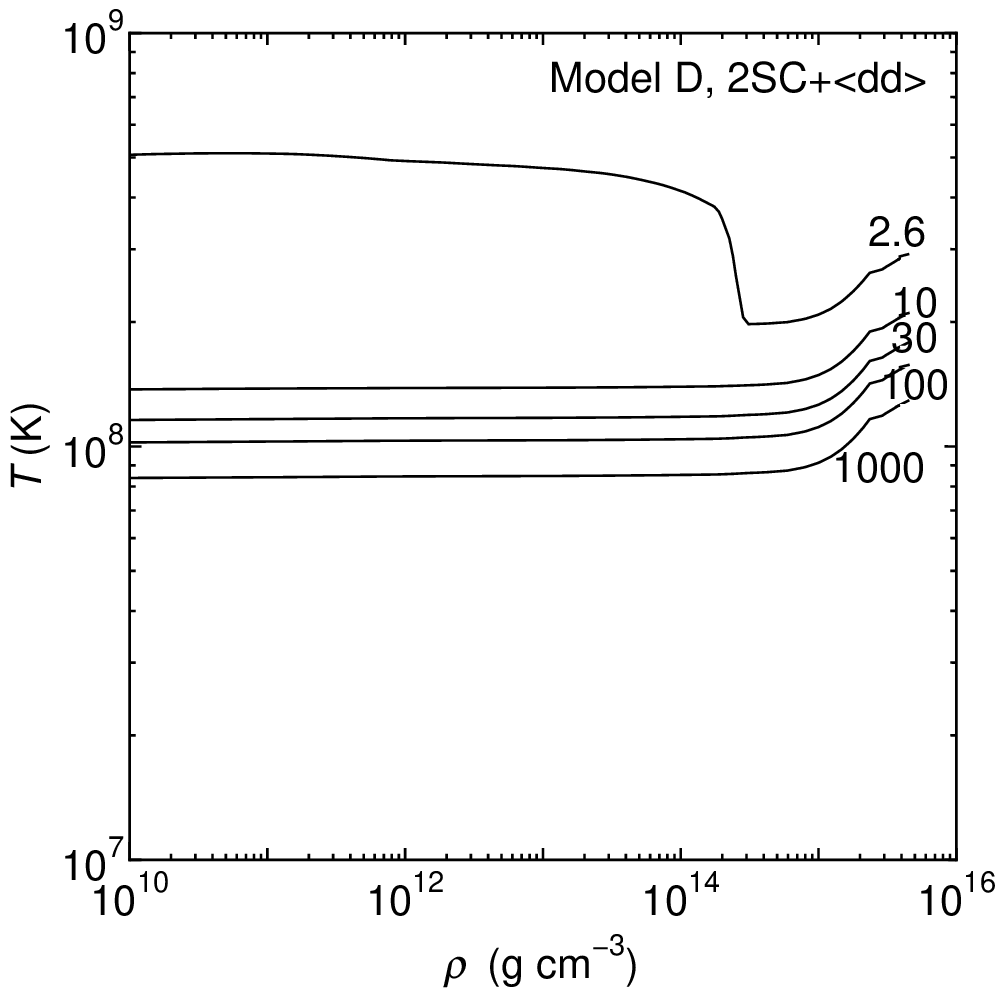}
\end{minipage}
\caption{Time evolution of temperature structure with superfluid model A (top) and D (bottom). The quark CSC is the only 2SC (left), the only CFL (middle), and 2SC+$\braket{dd}$ (right). The numerals attached to the curves show the ages (yr).}
\label{fig:strc}
\end{figure*}

In cooling curves with $2.13~M_{\odot}$ NSs, the surface temperature is much lower than the standard cooling models with 1.4 and 2 $M_{\odot}$ NSs. In particular, the models with the 2SC phase show supra-rapid cooling compared with CFL and 2SC+$\braket{dd}$ phases, which show similar cooling curves. Thus, we can distinguish 2SC and 2SC+$\braket{dd}$ from temperature observations. We also find that the ${}^3P_2$ superfluidity makes NSs warm with these models.

In model A (the maximum critical temperature with neutrons ${}^3P_2$ state of $\mathrm{max}(T^{n3}_{cr})=2\times10^{8}~{\rm K}$), cooling curves are located in lower temperature regions than all temperature observations, with any kinds of CSC. In the case of Model C ($\mathrm{max}(T^{n3}_{cr})=5\times10^{8}~{\rm K}$), three new observations of 3C58, Vela Jr., and Vela-like pulsar are passed through a cooling curve with CFL
and 2SC+$\braket{dd}$, not 2SC phase. Thus, the 2SC$+\braket{dd}$ phase is observational better than the only 2SC. Although Vela is higher than all the cooling curves, it would be explained with either lower-mass NSs or lighter surface compositions, which tend to be warmer.

Model D ($\mathrm{max}(T^{n3}_{cr})=1\times10^{9}~{\rm K}$) can explain the observation data of Vela by considering the 2SC+$\braket{dd}$ (or CFL) phase, but it cannot explain the lower limit of the observation with the 2SC phase.
But recently updated three observations are still too cold to be explained due to the large suppression of rapid cooling processes even with 2SC$+\braket{dd}$ or CFL phase. As the maximum mass with our EOS is $\approx2.13~M_{\odot}$, we suggest that strong 
neutrons ${}^3P_2$ superfluidity with $\mathrm{max}(T^{n3}_{cr})\gtrsim10^{9}~{\rm K}$ is not plausible. Considering that very weak neutron superfluidity such as Model A results in too rapid cooling compared to the observations as well, the suitable superfluid model is like Model C ($\mathrm{max}(T^{n3}_{cr})=5\times10^{8}~{\rm K}$) or slightly strong neutrons ${}^3P_2$ superfluid models. Coincidentally, this result is consistent with the constraint of
$\mathrm{max}(T^{n3}_{cr})$ obtained from the observed cooling rates of Cassiopeia A~\cite{2023MNRAS.518.2775S}.

For our best cooling model with Model C, we show the comparison of cooling curves between pure Fe and He envelope in FIG~\ref{fig:cche}. If there are light elements in the NS envelope, cooling curves tend to be located in higher temperature regions because the thermal conductivity of light elements is lower than that of Fe. This can be seen in both the slow cooling case ($M_{\rm NS}=1.41~M_{\odot}$) and the fast cooling case ($2.13~M_{\odot}$). By changing surface compositions, the consistency with the observations is changed. For example, $2.13~M_{\odot}$ NSs with pure Fe surface compositions are colder than the observed surface temperature of Vela, but the relationship becomes the opposite for the pure He surface compositions. This implies that, if Model C is true as the ${}^3P_2$ superfluid model, the surface compositions of Vela are intermediate between He and Fe. Thus, the uncertainties of surface compositions could alter the appropriate ${}^3P_2$ superfluid model.

FIG.~\ref{fig:strc} shows the internal temperature with models A and D. Let us consider model A, where the ${}^3P_2$ superfluidity is weakest in our models. At $t=2.6~{\rm yr}$, the NS core cools rapidly due to the presence of the DU process. Although the quark $\beta$ decay is active with the 2SC and 2SC+$\braket{dd}$ phase, the temperature structure is similar among different CSC states. The difference in CSC appears after the realization of isothermal structure for $t\gtrsim10~{\rm yr}$: The temperature with the 2SC phase rapidly decreases compared to the models with the 2SC+$\braket{dd}$ and CFL phases. It should be mentioned that even the weakest ${}^3P_2$ superfluidity alters the temperature structure when we compare it with the 2SC and 2SC +$\braket{dd}$ phases.

If the ${}^3P_2$ superfluid model is changed from A to D, you can see that the temperature becomes higher with the CFL and 2SC+$\braket{dd}$ phases, but cannot see so much with the 2SC phase. This is because the quark $\beta$ decay, which is derived from $d$ quarks in the  ${}^3P_2$ state, is suppressed except in the 2SC phase. Thus, we suggest that the ${}^3P_2$ superfluidity plays an important role in the 2SC+$\braket{dd}$ scenario. 

Considering CSC, there is an issue. The CFL paring is suitable for the cooling results which show moderately strong cooling (due to the DU process), but the appearing density is thought to
be much higher density. On the other hand, the 2SC pairing is suitable for the appearing density, but the cooling result shows too strong cooling to explain the observation. The 2SC+$\braket{dd}$ phase can solve this problem, as already suggested by Ref.~\cite{2005PhRvC..71d5801G} in the 2SC+X scenario. The appearing density of 2SC+$\braket{dd}$ is suitable for the realistic density of NS, and it \textit{can} suppress strong neutrino emission by quarks in all colours.

\section{Conclusion}
\label{sec:conc}

We investigate the thermal evolution
of hybrid stars with 2SC+$\braket{dd}$ phase which is expected from a new picture of quark-hadron continuity proposed by Ref.~\cite{2020PhRvD.101i4009F}. We found that cooling curves are drastically changed between 2SC and 2SC$+\braket{dd}$ phases, the latter of which is close to the CFL phase. Thus, we can clarify the difference between the 2SC and 2SC$+\braket{dd}$ phases through temperature observations such as Vela. 

We also found that ${}^{3}P_2$ superfluidity, associated with neutrons as well as $d$ quarks, plays an important role in the thermal evolution of NS in the case of the 2SC+$\braket{dd}$ as well as the CFL phase. This is a crucial difference compared to the only 2SC phase, where the ${}^{3}P_2$ superfluidity is not so important due to unsuppressed quark $\beta$ decay. If the (maximum) ${}^{3}P_2$ superfluid temperature lies in $5\times10^8~{\rm K}\lesssim \mathrm{max}(T^{n3}_{cr})<10^{9}~{\rm K}$, the 2SC+$\braket{dd}$ phase can account for the recent temperature observation of cold NSs such as Vela, 3C58, Vela Jr., and Vela-like pulsar, and therefore it is favored from temperature observations compared to the 2SC.

We adopted the hybrid EOS where the quark matter is constructed with the Dyson-Schwinger approach, which does not consider the di-quark condensation needed for the description of CSC. In regard to this, one of the concerns is that the mass-radius relationship may be changed. In the scenario with the 2SC+X phase, there is a kink structure to make the radius large at some masses regions. Since this can be regarded as the higher slope of the symmetry energy~\cite{2001ApJ...550..426L}, the proton fraction around the kink regions may be increased~\cite{2021PhRvD.104f3036K} and the DU process can be allowed in lower-mass NSs. However, our EOS also allows for the DU process, and the change from cooling curves in this paper is the only threshold mass of the DU process. 

However, if di-quark condensation is incorporated in the EOS of quark matter, the appearance density of 2SC+$\braket{dd}$ must be changed. The transition between 2SC+$\braket{dd}$ and CFL is also highly possible, though we did not consider such a case. The consistent cooling calculation with the EOS in terms of the CSC (and baryon superfluidity) is left for future work. Although there are many assumptions about nuclear models, this work is the first cooling calculation to consider quark-hadron continuity through the \textit{unified} ${}^{3}P_2$ superfluid model. Thus, we hope that this work provides the basis to pursue the evidence of another CSC state beyond the standard picture of the 2SC and CFL phases.

Recently-updated observations of cold NSs (3C58, Vela Jr., and Vela-like pulsar) beyond the minimal cooling scenario~\cite{2024arXiv240405371M} offer us a strong motivation to probe the exotic states in high-density matter. Our study is the first work to explore exotic states inside NSs with these updated observations. In particular, as fast neutrino cooling must be suppressed by the pairing for their observations, even with other EOSs, we believe it common that some kinds of superfluidity inside the NS core should not be too strong, such as the ${}^3P_2$ neutron superfluidity considered here. The desired information is their masses, which are little known (though they must be heavy due to the rapid cooling). If the mass measurement is fortunately performed in the future, with such as NICER, internal states (compositions and superfluidity/superconductivity) in NSs will be well determined. As one of the ingredients, the presence or absence of 2SC$+\braket{dd}$ examined in our work may possibly be judged.

In this study, we have assumed two major premises. One is that the EOS takes into account quark-hadron continuity by concerning first-order phase transitions; the other is that the critical temperature of the neutron's ${}^3P_2$ state is continuously linked to that of the $d$-quark. The former issue concerns the difficulty in defining the particle number fraction required for cooling calculations when considering the crossover EOS, whilst the latter concerns the excessive uncertainty surrounding the realistic ${}^3P_2$ critical temperature of the $d$-quark. We hope that in the near future we will be able to set aside these assumptions and engage in a self-consistent discussion.

\section*{Funding}
T.N. wishes to acknowledge the support from the Discretionary Budget of the President of Kurume Institute of Technology. This work is also supported by JSPS KAKENHI Grant Numbers JP23K19056, JP25K17403 (A.D.), JP24K07054, (N.Y.) and JP26K07094 (T.N.). 
\section*{Data Availability}
The rest of the data are available from the corresponding authors upon reasonable request.
\section*{Acknowledgments}
We thank T. Hatsuda for an informative discussion and Y. Fujimoto for careful reading of our manuscript. We thank the referee for the constructive comments, which helped improve the manuscript.

\bibliography{ref}{}

@ARTICLE{2011RPPh...74a4001F,
       author = {{Fukushima}, Kenji and {Hatsuda}, Tetsuo},
        title = "{The phase diagram of dense QCD}",
      journal = {Reports on Progress in Physics},
         year = 2011,
        month = jan,
       volume = {74},
       number = {1},
          eid = {014001},
        pages = {014001},
          doi = {10.1088/0034-4885/74/1/014001},
archivePrefix = {arXiv},
       eprint = {1005.4814},
 primaryClass = {hep-ph},
       adsurl = {https://ui.adsabs.harvard.edu/abs/2011RPPh...74a4001F}
}

@ARTICLE{1998PhLB..422..247A,
       author = {{Alford}, Mark and {Rajagopal}, Krishna and {Wilczek}, Frank},
        title = "{QCD at finite baryon density: nucleon droplets and color superconductivity}",
      journal = {Physics Letters B},
         year = 1998,
        month = mar,
       volume = {422},
       number = {1-4},
        pages = {247-256},
          doi = {10.1016/S0370-2693(98)00051-3},
archivePrefix = {arXiv},
       eprint = {hep-ph/9711395},
 primaryClass = {hep-ph},
       adsurl = {https://ui.adsabs.harvard.edu/abs/1998PhLB..422..247A}
}

@ARTICLE{1999NuPhB.537..443A,
       author = {{Alford}, Mark and {Rajagopal}, Krishna and {Wilczek}, Frank},
        title = "{Color-flavor locking and chiral symmetry breaking in high density QCD}",
      journal = {Nuclear Physics B},
         year = 1999,
        month = jan,
       volume = {537},
       number = {1},
        pages = {443-458},
          doi = {10.1016/S0550-3213(98)00668-3},
archivePrefix = {arXiv},
       eprint = {hep-ph/9804403},
 primaryClass = {hep-ph},
       adsurl = {https://ui.adsabs.harvard.edu/abs/1999NuPhB.537..443A}
}

@ARTICLE{1984PhR...107..325B,
       author = {{Bailin}, D. and {Love}, A.},
        title = "{Superfluidity and superconductivity in relativistic fermion systems}",
      journal = {\physrep},
         year = 1984,
        month = may,
       volume = {107},
       number = {6},
        pages = {325-385},
          doi = {10.1016/0370-1573(84)90145-5},
       adsurl = {https://ui.adsabs.harvard.edu/abs/1984PhR...107..325B}
}

@ARTICLE{2003NuPhA.714..481N,
       author = {{Neumann}, F. and {Buballa}, M. and {Oertel}, M.},
        title = "{Mixed phases of color superconducting quark matter}",
      journal = {\nphysa},
         year = 2003,
        month = feb,
       volume = {714},
       number = {3},
        pages = {481-501},
          doi = {10.1016/S0375-9474(02)01371-4},
archivePrefix = {arXiv},
       eprint = {hep-ph/0210078},
 primaryClass = {hep-ph},
       adsurl = {https://ui.adsabs.harvard.edu/abs/2003NuPhA.714..481N}
}

@ARTICLE{2002PhRvD..66i4007S,
       author = {{Steiner}, Andrew W. and {Reddy}, Sanjay and {Prakash}, Madappa},
        title = "{Color-neutral superconducting quark matter}",
      journal = {\prd},
         year = 2002,
        month = nov,
       volume = {66},
       number = {9},
          eid = {094007},
        pages = {094007},
          doi = {10.1103/PhysRevD.66.094007},
archivePrefix = {arXiv},
       eprint = {hep-ph/0205201},
 primaryClass = {hep-ph},
       adsurl = {https://ui.adsabs.harvard.edu/abs/2002PhRvD..66i4007S}
}

@ARTICLE{1970PThPh..44..905T,
       author = {{Tamagaki}, R.},
        title = "{Superfluid state in neutron star matter. I. Generalized Bogoliubov transformation and existence of $^{3}$P$_{2}$ gap at high density.}",
      journal = {Progress of Theoretical Physics},
         year = 1970,
        month = jan,
       volume = {44},
        pages = {905-928},
          doi = {10.1143/PTP.44.905},
       adsurl = {https://ui.adsabs.harvard.edu/abs/1970PThPh..44..905T}
}

@ARTICLE{2013arXiv1302.6626P,
      author = {{Page}, Dany and {Lattimer}, James M. and {Prakash}, Madappa and
         {Steiner}, Andrew W.},
        title = "{Stellar Superfluids}",
      journal = {Eds. K. H. Bennemann and J. B. Ketterson,},
         year = 2013,
        month = feb,
         booktitle = {"Novel Superfluids"},
         series = {Oxford University Press},
        pages = {505},
archivePrefix = {arXiv},
      eprint = {1302.6626},
 primaryClass = {astro-ph.HE},
      adsurl = {https://ui.adsabs.harvard.edu/abs/2013arXiv1302.6626P}
}

@ARTICLE{1993PThPS.112...27T,
       author = {{Takatsuka}, T. and {Tamagaki}, R.},
        title = "{Chapter II. Superfluidity in Neutron Star Matter and Symmetric Nuclear Matter}",
      journal = {Progress of Theoretical Physics Supplement},
         year = 1993,
        month = jan,
       volume = {112},
        pages = {27-65},
          doi = {10.1143/PTP.112.27},
       adsurl = {https://ui.adsabs.harvard.edu/abs/1993PThPS.112...27T}
}

@ARTICLE{1993NuPhA.555..128W,
       author = {{Wambach}, J. and {Ainsworth}, T.~L. and {Pines}, D.},
        title = "{Quasiparticle interactions in neutron matter for applications in neutron stars}",
      journal = {\nphysa},
         year = 1993,
        month = apr,
       volume = {555},
       number = {1},
        pages = {128-150},
          doi = {10.1016/0375-9474(93)90317-Q},
       adsurl = {https://ui.adsabs.harvard.edu/abs/1993NuPhA.555..128W}
}

@ARTICLE{2020PhRvD.101i4009F,
       author = {{Fujimoto}, Yuki and {Fukushima}, Kenji and {Weise}, Wolfram},
        title = "{Continuity from neutron matter to two-flavor quark matter with S$_{0}$ 1 and P$_{2}$ 3 superfluidity}",
      journal = {\prd},
         year = 2020,
        month = may,
       volume = {101},
       number = {9},
          eid = {094009},
        pages = {094009},
          doi = {10.1103/PhysRevD.101.094009},
archivePrefix = {arXiv},
       eprint = {1908.09360},
 primaryClass = {hep-ph},
       adsurl = {https://ui.adsabs.harvard.edu/abs/2020PhRvD.101i4009F}
}

@ARTICLE{2021PhRvD.104f3036K,
       author = {{Kojo}, Toru and {Hou}, Defu and {Okafor}, Jude and {Togashi}, Hajime},
        title = "{Phenomenological QCD equations of state for neutron star dynamics: Nuclear-2SC continuity and evolving effective couplings}",
      journal = {\prd},
         year = 2021,
        month = sep,
       volume = {104},
       number = {6},
          eid = {063036},
        pages = {063036},
          doi = {10.1103/PhysRevD.104.063036},
archivePrefix = {arXiv},
       eprint = {2012.01650},
 primaryClass = {astro-ph.HE},
       adsurl = {https://ui.adsabs.harvard.edu/abs/2021PhRvD.104f3036K}
}

@ARTICLE{2004ARA&A..42..169Y,
       author = {{Yakovlev}, D.~G. and {Pethick}, C.~J.},
        title = "{Neutron Star Cooling}",
      journal = {\araa},
         year = 2004,
        month = sep,
       volume = {42},
       number = {1},
        pages = {169-210},
          doi = {10.1146/annurev.astro.42.053102.134013},
archivePrefix = {arXiv},
       eprint = {astro-ph/0402143},
 primaryClass = {astro-ph},
       adsurl = {https://ui.adsabs.harvard.edu/abs/2004ARA&A..42..169Y}
}

@ARTICLE{2006NuPhA.777..497P,
       author = {{Page}, Dany and {Geppert}, Ulrich and {Weber}, Fridolin},
        title = "{The cooling of compact stars}",
      journal = {\nphysa},
         year = 2006,
        month = oct,
       volume = {777},
        pages = {497-530},
          doi = {10.1016/j.nuclphysa.2005.09.019},
archivePrefix = {arXiv},
       eprint = {astro-ph/0508056},
 primaryClass = {astro-ph},
       adsurl = {https://ui.adsabs.harvard.edu/abs/2006NuPhA.777..497P}
}

@ARTICLE{2020IJMPE..2930007K,
       author = {{Kim}, Myungkuk and {Lee}, Chang-Hwan and {Kim}, Young-Min and {Kwak}, Kyujin and {Lim}, Yeunhwan and {Hyun}, Chang Ho},
        title = "{Neutron star equations of state and their applications}",
      journal = {International Journal of Modern Physics E},
         year = 2020,
        month = jan,
       volume = {29},
       number = {7},
          eid = {2030007},
        pages = {2030007},
          doi = {10.1142/S0218301320300076},
       adsurl = {https://ui.adsabs.harvard.edu/abs/2020IJMPE..2930007K}
}

@ARTICLE{2022ApJ...934L..17R,
       author = {{Romani}, Roger W. and {Kandel}, D. and {Filippenko}, Alexei V. and {Brink}, Thomas G. and {Zheng}, WeiKang},
        title = "{PSR J0952-0607: The Fastest and Heaviest Known Galactic Neutron Star}",
      journal = {\apjl},
         year = 2022,
        month = aug,
       volume = {934},
       number = {2},
          eid = {L17},
        pages = {L17},
          doi = {10.3847/2041-8213/ac8007},
archivePrefix = {arXiv},
       eprint = {2207.05124},
 primaryClass = {astro-ph.HE},
       adsurl = {https://ui.adsabs.harvard.edu/abs/2022ApJ...934L..17R}
}

@ARTICLE{2021PrPNP.12003879B,
       author = {{Burgio}, G.~F. and {Schulze}, H. -J. and {Vida{\~n}a}, I. and {Wei}, J. -B.},
        title = "{Neutron stars and the nuclear equation of state}",
      journal = {Progress in Particle and Nuclear Physics},
         year = 2021,
        month = sep,
       volume = {120},
          eid = {103879},
        pages = {103879},
          doi = {10.1016/j.ppnp.2021.103879},
archivePrefix = {arXiv},
       eprint = {2105.03747},
 primaryClass = {nucl-th},
       adsurl = {https://ui.adsabs.harvard.edu/abs/2021PrPNP.12003879B}
}

@ARTICLE{2015PhRvC..91a5806H,
       author = {{Ho}, Wynn C.~G. and {Elshamouty}, Khaled G. and {Heinke}, Craig O. and {Potekhin}, Alexander Y.},
        title = "{Tests of the nuclear equation of state and superfluid and superconducting gaps using the Cassiopeia A neutron star}",
      journal = {\prc},
         year = 2015,
        month = jan,
       volume = {91},
       number = {1},
          eid = {015806},
        pages = {015806},
          doi = {10.1103/PhysRevC.91.015806},
archivePrefix = {arXiv},
       eprint = {1412.7759},
 primaryClass = {astro-ph.HE},
       adsurl = {https://ui.adsabs.harvard.edu/abs/2015PhRvC..91a5806H}
}

@ARTICLE{2003PhRvD..67e4018A,
       author = {{Alford}, Mark G. and {Bowers}, Jeffrey A. and {Cheyne}, Jack M. and {Cowan}, Greig A.},
        title = "{Single color and single flavor color superconductivity}",
      journal = {\prd},
         year = 2003,
        month = mar,
       volume = {67},
       number = {5},
          eid = {054018},
        pages = {054018},
          doi = {10.1103/PhysRevD.67.054018},
archivePrefix = {arXiv},
       eprint = {hep-ph/0210106},
 primaryClass = {hep-ph},
       adsurl = {https://ui.adsabs.harvard.edu/abs/2003PhRvD..67e4018A}
}

@ARTICLE{2011PhRvL.106h1101P,
       author = {{Page}, Dany and {Prakash}, Madappa and {Lattimer}, James M. and {Steiner}, Andrew W.},
        title = "{Rapid Cooling of the Neutron Star in Cassiopeia A Triggered by Neutron Superfluidity in Dense Matter}",
      journal = {\prl},
         year = 2011,
        month = feb,
       volume = {106},
       number = {8},
          eid = {081101},
        pages = {081101},
          doi = {10.1103/PhysRevLett.106.081101},
archivePrefix = {arXiv},
       eprint = {1011.6142},
 primaryClass = {astro-ph.HE},
       adsurl = {https://ui.adsabs.harvard.edu/abs/2011PhRvL.106h1101P}
}

@ARTICLE{2011MNRAS.412L.108S,
       author = {{Shternin}, Peter S. and {Yakovlev}, Dmitry G. and {Heinke}, Craig O. and {Ho}, Wynn C.~G. and {Patnaude}, Daniel J.},
        title = "{Cooling neutron star in the Cassiopeia A supernova remnant: evidence for superfluidity in the core}",
      journal = {\mnras},
         year = 2011,
        month = mar,
       volume = {412},
       number = {1},
        pages = {L108-L112},
          doi = {10.1111/j.1745-3933.2011.01015.x},
archivePrefix = {arXiv},
       eprint = {1012.0045},
 primaryClass = {astro-ph.SR},
       adsurl = {https://ui.adsabs.harvard.edu/abs/2011MNRAS.412L.108S}
}

@ARTICLE{2023MNRAS.518.2775S,
       author = {{Shternin}, Peter S. and {Ofengeim}, Dmitry D. and {Heinke}, Craig O. and {Ho}, Wynn C.~G.},
        title = "{Constraints on neutron star superfluidity from the cooling neutron star in Cassiopeia A using all Chandra ACIS-S observations}",
      journal = {\mnras},
         year = 2023,
        month = jan,
       volume = {518},
       number = {2},
        pages = {2775-2793},
          doi = {10.1093/mnras/stac3226},
archivePrefix = {arXiv},
       eprint = {2211.02526},
 primaryClass = {astro-ph.HE},
       adsurl = {https://ui.adsabs.harvard.edu/abs/2023MNRAS.518.2775S}
}

@ARTICLE{2000PhRvL..85.2048P,
       author = {{Page}, Dany and {Prakash}, Madappa and {Lattimer}, James M. and {Steiner}, Andrew W.},
        title = "{Prospects of Detecting Baryon and Quark Superfluidity from Cooling Neutron Stars}",
      journal = {\prl},
         year = 2000,
        month = sep,
       volume = {85},
       number = {10},
        pages = {2048-2051},
          doi = {10.1103/PhysRevLett.85.2048},
archivePrefix = {arXiv},
       eprint = {hep-ph/0005094},
 primaryClass = {hep-ph},
       adsurl = {https://ui.adsabs.harvard.edu/abs/2000PhRvL..85.2048P}
}

@ARTICLE{2000ApJ...533..406B,
       author = {{Blaschke}, D. and {Kl{\"a}hn}, T. and {Voskresensky}, D.~N.},
        title = "{Diquark Condensates and Compact Star Cooling}",
      journal = {\apj},
         year = 2000,
        month = apr,
       volume = {533},
       number = {1},
        pages = {406-412},
          doi = {10.1086/308664},
archivePrefix = {arXiv},
       eprint = {astro-ph/9908334},
 primaryClass = {astro-ph},
       adsurl = {https://ui.adsabs.harvard.edu/abs/2000ApJ...533..406B}
}

@ARTICLE{2001A&A...368..561B,
       author = {{Blaschke}, D. and {Grigorian}, H. and {Voskresensky}, D.~N.},
        title = "{Cooling of hybrid neutron stars and hypothetical self-bound objects with superconducting quark cores}",
      journal = {\aap},
         year = 2001,
        month = mar,
       volume = {368},
        pages = {561-568},
          doi = {10.1051/0004-6361:20010005},
archivePrefix = {arXiv},
       eprint = {astro-ph/0009120},
 primaryClass = {astro-ph},
       adsurl = {https://ui.adsabs.harvard.edu/abs/2001A&A...368..561B}
}

@ARTICLE{2005PhRvC..71d5801G,
       author = {{Grigorian}, Hovik and {Blaschke}, David and {Voskresensky}, Dmitri},
        title = "{Cooling of neutron stars with color superconducting quark cores}",
      journal = {\prc},
         year = 2005,
        month = apr,
       volume = {71},
       number = {4},
          eid = {045801},
        pages = {045801},
          doi = {10.1103/PhysRevC.71.045801},
archivePrefix = {arXiv},
       eprint = {astro-ph/0411619},
 primaryClass = {astro-ph},
       adsurl = {https://ui.adsabs.harvard.edu/abs/2005PhRvC..71d5801G}
}

@ARTICLE{2013ApJ...765....1N,
       author = {{Noda}, Tsuneo and {Hashimoto}, Masa-aki and {Yasutake}, Nobutoshi and {Maruyama}, Toshiki and {Tatsumi}, Toshitaka and {Fujimoto}, Masayuki},
        title = "{Cooling of Compact Stars with Color Superconducting Phase in Quark-hadron Mixed Phase}",
      journal = {\apj},
         year = 2013,
        month = mar,
       volume = {765},
       number = {1},
          eid = {1},
        pages = {1},
          doi = {10.1088/0004-637X/765/1/1},
archivePrefix = {arXiv},
       eprint = {1109.1080},
 primaryClass = {astro-ph.SR},
       adsurl = {https://ui.adsabs.harvard.edu/abs/2013ApJ...765....1N}
}

@ARTICLE{2016EPJA...52...44S,
       author = {{Sedrakian}, Armen},
        title = "{Cooling compact stars and phase transitions in dense QCD}",
      journal = {European Physical Journal A},
         year = 2016,
        month = mar,
       volume = {52},
          eid = {44},
        pages = {44},
          doi = {10.1140/epja/i2016-16044-y},
archivePrefix = {arXiv},
       eprint = {1509.06986},
 primaryClass = {astro-ph.HE},
       adsurl = {https://ui.adsabs.harvard.edu/abs/2016EPJA...52...44S}
}

@ARTICLE{2019EPJA...55..167S,
       author = {{Sedrakian}, Armen and {Clark}, John W.},
        title = "{Superfluidity in nuclear systems and neutron stars}",
      journal = {European Physical Journal A},
         year = 2019,
        month = sep,
       volume = {55},
       number = {9},
          eid = {167},
        pages = {167},
          doi = {10.1140/epja/i2019-12863-6},
archivePrefix = {arXiv},
       eprint = {1802.00017},
 primaryClass = {nucl-th},
       adsurl = {https://ui.adsabs.harvard.edu/abs/2019EPJA...55..167S}
}

@ARTICLE{2019PTEP.2019k3E01D,
      author = {{Dohi}, Akira and {Nakazato}, Ken'ichiro and {Hashimoto}, Masa-aki and
         {Yasuhide}, Matsuo and {Noda}, Tsuneo},
        title = "{Possibility of rapid neutron star cooling with a realistic equation of state}",
      journal = {\ptep},
         year = 2019,
        month = nov,
      volume = {2019},
      number = {11},
          eid = {113E01},
        pages = {113E01},
          doi = {10.1093/ptep/ptz116},
archivePrefix = {arXiv},
      eprint = {1910.01431},
 primaryClass = {astro-ph.HE},
      adsurl = {https://ui.adsabs.harvard.edu/abs/2019PTEP.2019k3E01D}
}

@ARTICLE{2022IJMPE..3150006D,
       author = {{Dohi}, Akira and {Liu}, Helei and {Noda}, Tsuneo and {Hashimoto}, Masa-Aki},
        title = "{Cooling of isolated neutron stars with pion condensation: Possible fast cooling in a low-symmetry energy model}",
      journal = {International Journal of Modern Physics E},
         year = 2022,
        month = jan,
       volume = {31},
       number = {2},
          eid = {2250006},
        pages = {2250006},
          doi = {10.1142/S0218301322500069},
archivePrefix = {arXiv},
       eprint = {2112.13302},
 primaryClass = {astro-ph.HE},
       adsurl = {https://ui.adsabs.harvard.edu/abs/2022IJMPE..3150006D}
}

@ARTICLE{1995A&A...297..717Y,
       author = {{Yakovlev}, D.~G. and {Levenfish}, K.~P.},
        title = "{Modified URCA process in neutron star cores.}",
      journal = {\aap},
         year = 1995,
        month = may,
       volume = {297},
        pages = {717},
       adsurl = {https://ui.adsabs.harvard.edu/abs/1995A&A...297..717Y}
}

@ARTICLE{1999A&A...343..650Y,
       author = {{Yakovlev}, D.~G. and {Kaminker}, A.~D. and {Levenfish}, K.~P.},
        title = "{Neutrino emission due to Cooper pairing of nucleons in cooling neutron stars}",
      journal = {\aap},
         year = 1999,
        month = mar,
       volume = {343},
        pages = {650-660},
          doi = {10.48550/arXiv.astro-ph/9812366},
archivePrefix = {arXiv},
       eprint = {astro-ph/9812366},
 primaryClass = {astro-ph},
       adsurl = {https://ui.adsabs.harvard.edu/abs/1999A&A...343..650Y}
}

@ARTICLE{2020ApJ...898..125P,
       author = {{Page}, Dany and {Beznogov}, Mikhail V. and {Garibay}, Iv{\'a}n and {Lattimer}, James M. and {Prakash}, Madappa and {Janka}, Hans-Thomas},
        title = "{Ns 1987A in SN 1987A}",
      journal = {\apj},
         year = 2020,
        month = jul,
       volume = {898},
       number = {2},
          eid = {125},
        pages = {125},
          doi = {10.3847/1538-4357/ab93c2},
archivePrefix = {arXiv},
       eprint = {2004.06078},
 primaryClass = {astro-ph.HE},
       adsurl = {https://ui.adsabs.harvard.edu/abs/2020ApJ...898..125P}
}

@ARTICLE{2023ApJ...949...97D,
       author = {{Dohi}, Akira and {Greco}, Emanuele and {Nagataki}, Shigehiro and {Ono}, Masaomi and {Miceli}, Marco and {Orlando}, Salvatore and {Olmi}, Barbara},
        title = "{Investigating the Time Evolution of the Thermal Emission from the Putative Neutron Star in SN 1987A for 50+ Years}",
      journal = {\apj},
         year = 2023,
        month = jun,
       volume = {949},
       number = {2},
          eid = {97},
        pages = {97},
          doi = {10.3847/1538-4357/acce3f},
archivePrefix = {arXiv},
       eprint = {2304.08418},
 primaryClass = {astro-ph.HE},
       adsurl = {https://ui.adsabs.harvard.edu/abs/2023ApJ...949...97D}
}

@ARTICLE{1980PhRvL..44.1637I,
       author = {{Iwamoto}, N.},
        title = "{Quark Beta Decay and the Cooling of Neutron Stars}",
      journal = {\prl},
         year = 1980,
        month = jun,
       volume = {44},
       number = {24},
        pages = {1637-1640},
          doi = {10.1103/PhysRevLett.44.1637},
       adsurl = {https://ui.adsabs.harvard.edu/abs/1980PhRvL..44.1637I}
}

@ARTICLE{1982AnPhy.141....1I,
       author = {{Iwamoto}, N.},
        title = "{Neutrino emissivities and mean free paths of degenerate quark matter.}",
      journal = {Annals of Physics},
         year = 1982,
        month = jan,
       volume = {141},
        pages = {1-49},
          doi = {10.1016/0003-4916(82)90271-8},
       adsurl = {https://ui.adsabs.harvard.edu/abs/1982AnPhy.141....1I}
}

@ARTICLE{1999AcPPB..30.1125K,
       author = {{Kaminker}, A.~D. and {Haensel}, P.},
        title = "{Neutrino emission due to electron bremsstrahlung in superfluid neutron-star cores.}",
      journal = {Acta Physica Polonica B},
         year = 1999,
        month = apr,
       volume = {30},
       number = {4},
        pages = {1125-1148},
          doi = {10.48550/arXiv.astro-ph/9908249},
archivePrefix = {arXiv},
       eprint = {astro-ph/9908249},
 primaryClass = {astro-ph},
       adsurl = {https://ui.adsabs.harvard.edu/abs/1999AcPPB..30.1125K}
}

@INPROCEEDINGS{2016JPhCS.665a2068Y,
       author = {{Yasutake}, N. and {Chen}, H. and {Maruyama}, T. and {Tatsumi}, T.},
        title = "{Finite size effects in hadron-quark phase transition by the Dyson-Schwinger method}",
    booktitle = {Journal of Physics Conference Series},
         year = 2016,
       series = {Journal of Physics Conference Series},
       volume = {665},
        month = jan,
          eid = {012068},
        pages = {012068},
          doi = {10.1088/1742-6596/665/1/012068},
archivePrefix = {arXiv},
       eprint = {1309.1954},
 primaryClass = {astro-ph.HE},
       adsurl = {https://ui.adsabs.harvard.edu/abs/2016JPhCS.665a2068Y}
}

@ARTICLE{2009PhRvD..80l3009Y,
       author = {{Yasutake}, Nobutoshi and {Maruyama}, Toshiki and {Tatsumi}, Toshitaka},
        title = "{Hot hadron-quark mixed phase including hyperons}",
      journal = {\prd},
         year = 2009,
        month = dec,
       volume = {80},
       number = {12},
          eid = {123009},
        pages = {123009},
          doi = {10.1103/PhysRevD.80.123009},
archivePrefix = {arXiv},
       eprint = {0910.1144},
 primaryClass = {nucl-th},
       adsurl = {https://ui.adsabs.harvard.edu/abs/2009PhRvD..80l3009Y}
}

@ARTICLE{2012arXiv1208.0427Y,
       author = {{Yasutake}, Nobutoshi and {Noda}, Tsuneo and {Sotani}, Hajime and {Maruyama}, Toshiki and {Tatsumi}, Toshitaka},
        title = "{Thermodynamical description of hadron-quark phase transition and its implications on compact-star phenomena}",
      journal = {},
          year = 2013,
        month = feb,
         booktitle = {Recent Advances in Quarks Research},
         series = {Nova Science Publishers},
        pages = {63-112},
          doi = {10.48550/arXiv.1208.0427},
archivePrefix = {arXiv},
       eprint = {1208.0427},
 primaryClass = {astro-ph.HE},
       adsurl = {https://ui.adsabs.harvard.edu/abs/2012arXiv1208.0427Y}
}

@ARTICLE{2023arXiv230813973M,
       author = {{Mariani}, Mauro and {Lugones}, Germ{\'a}n},
        title = "{Quark-hadron pasta phase in neutron stars: the role of medium-dependent surface and curvature tensions}",
      journal = {arXiv e-prints},
         year = 2023,
        month = aug,
          eid = {arXiv:2308.13973},
        pages = {arXiv:2308.13973},
          doi = {10.48550/arXiv.2308.13973},
archivePrefix = {arXiv},
       eprint = {2308.13973},
 primaryClass = {nucl-th},
       adsurl = {https://ui.adsabs.harvard.edu/abs/2023arXiv230813973M}
}

@ARTICLE{2023PhRvD.107j3009Q,
       author = {{Qin}, Pianpian and {Bai}, Zhan and {Wang}, Sibo and {Wang}, Chencan and {Qin}, Si-xue},
        title = "{Hadron-quark phase transition in neutron star by combining the relativistic Brueckner-Hartree-Fock theory and Dyson-Schwinger equation approach}",
      journal = {\prd},
         year = 2023,
        month = may,
       volume = {107},
       number = {10},
          eid = {103009},
        pages = {103009},
          doi = {10.1103/PhysRevD.107.103009},
archivePrefix = {arXiv},
       eprint = {2301.02768},
 primaryClass = {nucl-th},
       adsurl = {https://ui.adsabs.harvard.edu/abs/2023PhRvD.107j3009Q}
}

@ARTICLE{1987PhR...149....1M,
       author = {{Machleidt}, R. and {Holinde}, K. and {Elster}, Ch.},
        title = "{The bonn meson-exchange model for the nucleon{\textemdash}nucleon interaction}",
      journal = {\physrep},
         year = 1987,
        month = jan,
       volume = {149},
       number = {1},
        pages = {1-89},
          doi = {10.1016/S0370-1573(87)80002-9},
       adsurl = {https://ui.adsabs.harvard.edu/abs/1987PhR...149....1M}
}

@ARTICLE{1997PhRvC..56.1720P,
       author = {{Pudliner}, B.~S. and {Pandharipande}, V.~R. and {Carlson}, J. and {Pieper}, Steven C. and {Wiringa}, R.~B.},
        title = "{Quantum Monte Carlo calculations of nuclei with A<=7}",
      journal = {\prc},
         year = 1997,
        month = oct,
       volume = {56},
       number = {4},
        pages = {1720-1750},
          doi = {10.1103/PhysRevC.56.1720},
archivePrefix = {arXiv},
       eprint = {nucl-th/9705009},
 primaryClass = {nucl-th},
       adsurl = {https://ui.adsabs.harvard.edu/abs/1997PhRvC..56.1720P}
}

@ARTICLE{2007PhRvD..76l3015M,
       author = {{Maruyama}, Toshiki and {Chiba}, Satoshi and {Schulze}, Hans-Josef and {Tatsumi}, Toshitaka},
        title = "{Hadron-quark mixed phase in hyperon stars}",
      journal = {\prd},
         year = 2007,
        month = dec,
       volume = {76},
       number = {12},
          eid = {123015},
        pages = {123015},
          doi = {10.1103/PhysRevD.76.123015},
archivePrefix = {arXiv},
       eprint = {0708.3277},
 primaryClass = {nucl-th},
       adsurl = {https://ui.adsabs.harvard.edu/abs/2007PhRvD..76l3015M}
}

@ARTICLE{2015ApJ...813..135M,
       author = {{Miyatsu}, Tsuyoshi and {Cheoun}, Myung-Ki and {Saito}, Koichi},
        title = "{Equation of State for Neutron Stars with Hyperons and Quarks in the Relativistic Hartree-Fock Approximation}",
      journal = {\apj},
         year = 2015,
        month = nov,
       volume = {813},
       number = {2},
          eid = {135},
        pages = {135},
          doi = {10.1088/0004-637X/813/2/135},
archivePrefix = {arXiv},
       eprint = {1506.05552},
 primaryClass = {nucl-th},
       adsurl = {https://ui.adsabs.harvard.edu/abs/2015ApJ...813..135M}
}

@ARTICLE{2018PhRvD..97b3018B,
       author = {{Bai}, Zhan and {Chen}, Huan and {Liu}, Yu-xin},
        title = "{Revisiting the equation of state of hybrid stars in the Dyson-Schwinger equation approach to QCD}",
      journal = {\prd},
         year = 2018,
        month = jan,
       volume = {97},
       number = {2},
          eid = {023018},
        pages = {023018},
          doi = {10.1103/PhysRevD.97.023018},
archivePrefix = {arXiv},
       eprint = {1707.09535},
 primaryClass = {hep-ph},
       adsurl = {https://ui.adsabs.harvard.edu/abs/2018PhRvD..97b3018B}
}

@ARTICLE{2006PThPh.115..355T,
       author = {{Takatsuka}, T. and {Nishizaki}, S. and {Yamamoto}, Y. and {Tamagaki}, R.},
        title = "{Occurrence of Hyperon Superfluidity in Neutron Star Cores}",
      journal = {Progress of Theoretical Physics},
         year = 2006,
        month = feb,
       volume = {115},
       number = {2},
        pages = {355-379},
          doi = {10.1143/PTP.115.355},
archivePrefix = {arXiv},
       eprint = {nucl-th/0601043},
 primaryClass = {nucl-th},
       adsurl = {https://ui.adsabs.harvard.edu/abs/2006PThPh.115..355T}
}

@ARTICLE{2018PhRvL.121p1101A,
      author = {{Abbott}, B.~P. and {Abbott}, R. and {Abbott}, T.~D. and {Acernese}, F. and
         {Ackley}, K. {\it et al.}, {LIGO Scientific Collaboration} and
         {Virgo Collaboration}},
        title = "{GW170817: Measurements of Neutron Star Radii and Equation of State}",
      journal = {\prl},
         year = 2018,
        month = oct,
      volume = {121},
      number = {16},
          eid = {161101},
        pages = {161101},
          doi = {10.1103/PhysRevLett.121.161101},
archivePrefix = {arXiv},
      eprint = {1805.11581},
 primaryClass = {gr-qc},
      adsurl = {https://ui.adsabs.harvard.edu/abs/2018PhRvL.121p1101A}
}

@ARTICLE{2002PhRvD..65i4026A,
       author = {{Alkofer}, R. and {Watson}, P. and {Weigel}, H.},
        title = "{Mesons in a Poincar{\'e} covariant Bethe-Salpeter approach}",
      journal = {\prd},
         year = 2002,
        month = may,
       volume = {65},
       number = {9},
          eid = {094026},
        pages = {094026},
          doi = {10.1103/PhysRevD.65.094026},
archivePrefix = {arXiv},
       eprint = {hep-ph/0202053},
 primaryClass = {hep-ph},
       adsurl = {https://ui.adsabs.harvard.edu/abs/2002PhRvD..65i4026A}
}

@ARTICLE{2013Sci...340..448A,
      author = {{Antoniadis}, John and {Freire}, Paulo C.~C. and {Wex}, Norbert and
         {Tauris}, Thomas M. and {Lynch {\it et al.}}, Ryan S.},
        title = "{A Massive Pulsar in a Compact Relativistic Binary}",
      journal = {Science},
         year = 2013,
        month = apr,
      volume = {340},
      number = {6131},
        pages = {448},
          doi = {10.1126/science.1233232},
archivePrefix = {arXiv},
      eprint = {1304.6875},
 primaryClass = {astro-ph.HE},
      adsurl = {https://ui.adsabs.harvard.edu/abs/2013Sci...340..448A}
}

@ARTICLE{2021Univ....7..182L,
       author = {{Li}, Bao-An and {Cai}, Bao-Jun and {Xie}, Wen-Jie and {Zhang}, Nai-Bo},
        title = "{Progress in Constraining Nuclear Symmetry Energy Using Neutron Star Observables Since GW170817}",
      journal = {Universe},
         year = 2021,
        month = jun,
       volume = {7},
       number = {6},
          eid = {182},
        pages = {182},
          doi = {10.3390/universe7060182},
archivePrefix = {arXiv},
       eprint = {2105.04629},
 primaryClass = {nucl-th},
       adsurl = {https://ui.adsabs.harvard.edu/abs/2021Univ....7..182L}
}

@ARTICLE{2011PhRvD..84j5023C,
       author = {{Chen}, H. and {Baldo}, M. and {Burgio}, G.~F. and {Schulze}, H. -J.},
        title = "{Hybrid stars with the Dyson-Schwinger quark model}",
      journal = {\prd},
         year = 2011,
        month = nov,
       volume = {84},
       number = {10},
          eid = {105023},
        pages = {105023},
          doi = {10.1103/PhysRevD.84.105023},
archivePrefix = {arXiv},
       eprint = {1107.2497},
 primaryClass = {nucl-th},
       adsurl = {https://ui.adsabs.harvard.edu/abs/2011PhRvD..84j5023C}
}

@ARTICLE{2020NatAs...4...72C,
       author = {{Cromartie}, H.~T. and {Fonseca}, E. and {Ransom}, S.~M. and
         {Demorest}, P.~B. and {Arzoumanian {\it et al.}}, Z.},
        title = "{Relativistic Shapiro delay measurements of an extremely massive millisecond pulsar}",
      journal = {Nat.~Astron.},
         year = 2020,
        month = jan,
       volume = {4},
        pages = {72-76},
          doi = {10.1038/s41550-019-0880-2},
archivePrefix = {arXiv},
       eprint = {1904.06759},
 primaryClass = {astro-ph.HE},
       adsurl = {https://ui.adsabs.harvard.edu/abs/2020NatAs...4...72C}
}

@ARTICLE{2010Natur.467.1081D,
      author = {{Demorest}, P.~B. and {Pennucci}, T. and {Ransom}, S.~M. and
         {Roberts}, M.~S.~E. and {Hessels}, J.~W.~T.},
        title = "{A two-solar-mass neutron star measured using Shapiro delay}",
      journal = {\nat},
         year = 2010,
        month = oct,
      volume = {467},
      number = {7319},
        pages = {1081-1083},
          doi = {10.1038/nature09466},
archivePrefix = {arXiv},
      eprint = {1010.5788},
 primaryClass = {astro-ph.HE},
      adsurl = {https://ui.adsabs.harvard.edu/abs/2010Natur.467.1081D}
}

@ARTICLE{1984ApJ...278..813F,
      author = {{Fujimoto}, M.~Y. and {Hanawa}, T. and {Iben}, I., Jr. and
         {Richardson}, M.~B.},
        title = "{Thermal evolution of accreting neutron stars}",
      journal = {\apj},
         year = 1984,
        month = mar,
      volume = {278},
        pages = {813-824},
          doi = {10.1086/161851},
      adsurl = {https://ui.adsabs.harvard.edu/abs/1984ApJ...278..813F}
}

@ARTICLE{2009ApJ...691.1035H,
      author = {{Heinke}, C.~O. and {Jonker}, P.~G. and {Wijnands}, R. and
         {Deloye}, C.~J. and {Taam}, R.~E.},
        title = "{Further Constraints on Thermal Quiescent X-Ray Emission from SAX J1808.4-3658}",
      journal = {\apj},
         year = 2009,
        month = feb,
      volume = {691},
      number = {2},
        pages = {1035-1041},
          doi = {10.1088/0004-637X/691/2/1035},
archivePrefix = {arXiv},
      eprint = {0810.0497},
 primaryClass = {astro-ph},
      adsurl = {https://ui.adsabs.harvard.edu/abs/2009ApJ...691.1035H}
}

@ARTICLE{2002PhRvD..66f3003J,
       author = {{Jaikumar}, Prashanth and {Prakash}, Madappa and {Sch{\"a}fer}, Thomas},
        title = "{Neutrino emission from Goldstone modes in dense quark matter}",
      journal = {\prd},
         year = 2002,
        month = sep,
       volume = {66},
       number = {6},
          eid = {063003},
        pages = {063003},
          doi = {10.1103/PhysRevD.66.063003},
archivePrefix = {arXiv},
       eprint = {astro-ph/0203088},
 primaryClass = {astro-ph},
       adsurl = {https://ui.adsabs.harvard.edu/abs/2002PhRvD..66f3003J}
}

@ARTICLE{2004ApJS..153..269K,
       author = {{Kaplan}, D.~L. and {Frail}, D.~A. and {Gaensler}, B.~M. and
         {Gotthelf}, E.~V. and {Kulkarni}, S.~R. and {Slane}, P.~O. and
         {Nechita}, A.},
        title = "{An X-Ray Search for Compact Central Sources in Supernova Remnants. I. SNRS G093.3+6.9, G315.4-2.3, G084.2+0.8, and G127.1+0.5}",
      journal = {\apjs},
         year = 2004,
        month = jul,
       volume = {153},
       number = {1},
        pages = {269-315},
          doi = {10.1086/421065},
archivePrefix = {arXiv},
       eprint = {astro-ph/0403313},
 primaryClass = {astro-ph},
       adsurl = {https://ui.adsabs.harvard.edu/abs/2004ApJS..153..269K}
}

@ARTICLE{2001ApJ...550..426L,
       author = {{Lattimer}, J.~M. and {Prakash}, M.},
        title = "{Neutron Star Structure and the Equation of State}",
      journal = {\apj},
         year = 2001,
        month = mar,
       volume = {550},
       number = {1},
        pages = {426-442},
          doi = {10.1086/319702},
archivePrefix = {arXiv},
       eprint = {astro-ph/0002232},
 primaryClass = {astro-ph},
       adsurl = {https://ui.adsabs.harvard.edu/abs/2001ApJ...550..426L}
}

@ARTICLE{2017IJMPE..2650015L,
      author = {{Lim}, Yeunhwan and {Hyun}, Chang Ho and {Lee}, Chang-Hwan},
        title = "{Nuclear equation of state and neutron star cooling}",
      journal = {Int.~J.~Mod.~Phys.~E},
         year = 2017,
        month = jan,
      volume = {26},
      number = {4},
          eid = {1750015-328},
        pages = {1750015-328},
          doi = {10.1142/S021830131750015X},
      adsurl = {https://ui.adsabs.harvard.edu/abs/2017IJMPE..2650015L}
}

@ARTICLE{2021ApJ...918L..28M,
       author = {{Miller}, M.~C. and {Lamb}, F.~K. and {Dittmann}, A.~J. and {Bogdanov}, S. and {Arzoumanian}, Z. and {Gendreau}, K.~C. and {Guillot}, S. and {Ho}, W.~C.~G. and {Lattimer}, J.~M. and {Loewenstein}, M. and {Morsink}, S.~M. and {Ray}, P.~S. and {Wolff}, M.~T. and {Baker}, C.~L. and {Cazeau}, T. and {Manthripragada}, S. and {Markwardt}, C.~B. and {Okajima}, T. and {Pollard}, S. and {Cognard}, I. and {Cromartie}, H.~T. and {Fonseca}, E. and {Guillemot}, L. and {Kerr}, M. and {Parthasarathy}, A. and {Pennucci}, T.~T. and {Ransom}, S. and {Stairs}, I.},
        title = "{The Radius of PSR J0740+6620 from NICER and XMM-Newton Data}",
      journal = {\apjl},
         year = 2021,
        month = sep,
       volume = {918},
       number = {2},
          eid = {L28},
        pages = {L28},
          doi = {10.3847/2041-8213/ac089b},
archivePrefix = {arXiv},
       eprint = {2105.06979},
 primaryClass = {astro-ph.HE},
       adsurl = {https://ui.adsabs.harvard.edu/abs/2021ApJ...918L..28M}
}

@ARTICLE{1939PhRv...55..374O,
      author = {{Oppenheimer}, J.~R. and {Volkoff}, G.~M.},
        title = "{On Massive Neutron Cores}",
      journal = {Physical Review},
         year = 1939,
        month = feb,
      volume = {55},
      number = {4},
        pages = {374-381},
          doi = {10.1103/PhysRev.55.374},
      adsurl = {https://ui.adsabs.harvard.edu/abs/1939PhRv...55..374O}
}

@ARTICLE{2004ApJS..155..623P,
      author = {{Page}, Dany and {Lattimer}, James M. and {Prakash}, Madappa and
         {Steiner}, Andrew W.},
        title = "{Minimal Cooling of Neutron Stars: A New Paradigm}",
      journal = {\apjs},
         year = 2004,
        month = dec,
      volume = {155},
      number = {2},
        pages = {623-650},
          doi = {10.1086/424844},
archivePrefix = {arXiv},
      eprint = {astro-ph/0403657},
 primaryClass = {astro-ph},
      adsurl = {https://ui.adsabs.harvard.edu/abs/2004ApJS..155..623P}
}

@ARTICLE{2018A&A...609A..74P,
      author = {{Potekhin}, A.~Y. and {Chabrier}, G.},
        title = "{Magnetic neutron star cooling and microphysics}",
      journal = {\aap},
         year = 2018,
        month = jan,
      volume = {609},
          eid = {A74},
        pages = {A74},
          doi = {10.1051/0004-6361/201731866},
archivePrefix = {arXiv},
      eprint = {1711.07662},
 primaryClass = {astro-ph.HE},
      adsurl = {https://ui.adsabs.harvard.edu/abs/2018A&A...609A..74P}
}

@ARTICLE{1992ApJ...390L..77P,
      author = {{Prakash}, Madappa and {Prakash}, Manju and {Lattimer}, James M. and
         {Pethick}, C.~J.},
        title = "{Rapid Cooling of Neutron Stars by Hyperons and Delta Isobars}",
      journal = {\apjl},
         year = 1992,
        month = may,
      volume = {390},
        pages = {L77},
          doi = {10.1086/186376},
      adsurl = {https://ui.adsabs.harvard.edu/abs/1992ApJ...390L..77P}
}

@ARTICLE{1994PrPNP..33..477R,
       author = {{Roberts}, Craig D. and {Williams}, Anthony G.},
        title = "{Dyson-Schwinger equations and their application to hadronic physics}",
      journal = {Progress in Particle and Nuclear Physics},
         year = 1994,
        month = jan,
       volume = {33},
        pages = {477-575},
          doi = {10.1016/0146-6410(94)90049-3},
archivePrefix = {arXiv},
       eprint = {hep-ph/9403224},
 primaryClass = {hep-ph},
       adsurl = {https://ui.adsabs.harvard.edu/abs/1994PrPNP..33..477R}
}

@ARTICLE{2003NuPhA.714..337R,
       author = {{Reddy}, Sanjay and {Sadzikowski}, Mariusz and {Tachibana}, Motoi},
        title = "{Neutrino rates in color flavor locked quark matter}",
      journal = {\nphysa},
         year = 2003,
        month = feb,
       volume = {714},
       number = {1-2},
        pages = {337-351},
          doi = {10.1016/S0375-9474(02)01351-9},
archivePrefix = {arXiv},
       eprint = {nucl-th/0203011},
 primaryClass = {astro-ph},
       adsurl = {https://ui.adsabs.harvard.edu/abs/2003NuPhA.714..337R}
}

@ARTICLE{2004PThPh.112...37T,
      author = {{Takatsuka}, T. and {Tamagaki}, R.},
        title = "{Baryon Superfluidity and Neutrino Emissivity of Neutron Stars}",
      journal = {\ptp},
         year = 2004,
        month = jul,
      volume = {112},
      number = {1},
        pages = {37-72},
          doi = {10.1143/PTP.112.37},
archivePrefix = {arXiv},
      eprint = {nucl-th/0402011},
 primaryClass = {nucl-th},
      adsurl = {https://ui.adsabs.harvard.edu/abs/2004PThPh.112...37T}
}

@ARTICLE{2020PrPNP.11203770T,
       author = {{Tolos}, L. and {Fabbietti}, L.},
        title = "{Strangeness in nuclei and neutron stars}",
      journal = {Progress in Particle and Nuclear Physics},
         year = 2020,
        month = may,
       volume = {112},
          eid = {103770},
        pages = {103770},
          doi = {10.1016/j.ppnp.2020.103770},
archivePrefix = {arXiv},
       eprint = {2002.09223},
 primaryClass = {nucl-ex},
       adsurl = {https://ui.adsabs.harvard.edu/abs/2020PrPNP.11203770T}
}

@ARTICLE{1998PhR...292....1T,
      author = {{Tsuruta}, S.},
        title = "{Thermal properties and detectability of neutron stars. II. Thermal evolution of rotation-powered neutron stars}",
      journal = {\physrep},
         year = 1998,
        month = jan,
      volume = {292},
        pages = {1-130},
          doi = {10.1016/S0370-1573(97)00041-0},
      adsurl = {https://ui.adsabs.harvard.edu/abs/1998PhR...292....1T}
}

@ARTICLE{1977ApJ...212..825T,
      author = {{Thorne}, K.~S.},
        title = "{The relativistic equations of stellar structure and evolution.}",
      journal = {\apj},
         year = 1977,
        month = mar,
      volume = {212},
        pages = {825-831},
          doi = {10.1086/155108},
      adsurl = {https://ui.adsabs.harvard.edu/abs/1977ApJ...212..825T}
}

@ARTICLE{1939PhRv...55..364T,
      author = {{Tolman}, Richard C.},
        title = "{Static Solutions of Einstein's Field Equations for Spheres of Fluid}",
      journal = {Physical Review},
         year = 1939,
        month = feb,
      volume = {55},
      number = {4},
        pages = {364-373},
          doi = {10.1103/PhysRev.55.364},
      adsurl = {https://ui.adsabs.harvard.edu/abs/1939PhRv...55..364T}
}

@ARTICLE{2001PhR...354....1Y,
      author = {{Yakovlev}, D.~G. and {Kaminker}, A.~D. and {Gnedin}, O.~Y. and
         {Haensel}, P.},
        title = "{Neutrino emission from neutron stars}",
      journal = {\physrep},
         year = 2001,
        month = nov,
      volume = {354},
      number = {1-2},
        pages = {1-155},
          doi = {10.1016/S0370-1573(00)00131-9},
archivePrefix = {arXiv},
      eprint = {astro-ph/0012122},
 primaryClass = {astro-ph},
      adsurl = {https://ui.adsabs.harvard.edu/abs/2001PhR...354....1Y}
}

@ARTICLE{2024arXiv240405371M,
       author = {{Marino}, Alessio and {Dehman}, Clara and {Kovlakas}, Konstantinos and {Rea}, Nanda and {Pons}, Jose A. and {Vigan{\`o}}, D.},
        title = "{Constraints on the dense matter equation of state from young and cold isolated neutron stars}",
      journal = {arXiv e-prints},
         year = 2024,
        month = apr,
          eid = {arXiv:2404.05371},
        pages = {arXiv:2404.05371},
          doi = {10.48550/arXiv.2404.05371},
archivePrefix = {arXiv},
       eprint = {2404.05371},
 primaryClass = {astro-ph.HE},
       adsurl = {https://ui.adsabs.harvard.edu/abs/2024arXiv240405371M}
}

@ARTICLE{2015ApJ...798...82A,
       author = {{Allen}, G.~E. and {Chow}, K. and {DeLaney}, T. and {Filipovi{\'c}}, M.~D. and {Houck}, J.~C. and {Pannuti}, T.~G. and {Stage}, M.~D.},
        title = "{On the Expansion Rate, Age, and Distance of the Supernova Remnant G266.2-1.2 (Vela Jr.)}",
      journal = {\apj},
         year = 2015,
        month = jan,
       volume = {798},
       number = {2},
          eid = {82},
        pages = {82},
          doi = {10.1088/0004-637X/798/2/82},
archivePrefix = {arXiv},
       eprint = {1410.7435},
 primaryClass = {astro-ph.HE},
       adsurl = {https://ui.adsabs.harvard.edu/abs/2015ApJ...798...82A}
}

@ARTICLE{2023A&A...673A..45C,
       author = {{Camilloni}, Francesco and {Becker}, Werner and {Predehl}, Peter and {Dennerl}, Konrad and {Freyberg}, Michael and {Mayer}, Martin G.~F. and {Sasaki}, Manami},
        title = "{SRG/eROSITA and XMM-Newton observations of Vela Jr}",
      journal = {\aap},
         year = 2023,
        month = may,
       volume = {673},
          eid = {A45},
        pages = {A45},
          doi = {10.1051/0004-6361/202245475},
archivePrefix = {arXiv},
       eprint = {2303.12686},
 primaryClass = {astro-ph.HE},
       adsurl = {https://ui.adsabs.harvard.edu/abs/2023A&A...673A..45C}
}

@ARTICLE{2013A&A...560A..18K,
       author = {{Kothes}, R.},
        title = "{Distance and age of the pulsar wind nebula 3C 58}",
      journal = {\aap},
         year = 2013,
        month = dec,
       volume = {560},
          eid = {A18},
        pages = {A18},
          doi = {10.1051/0004-6361/201219839},
archivePrefix = {arXiv},
       eprint = {1307.8384},
 primaryClass = {astro-ph.GA},
       adsurl = {https://ui.adsabs.harvard.edu/abs/2013A&A...560A..18K}
}

@INPROCEEDINGS{2022EPJWC.26011024N,
       author = {{Noda}, Tsuneo and {Yasutake}, Nobutoshi and {Hashimoto}, Masa-aki and {Maruyama}, Toshiki and {Tatsumi}, Toshitaka},
        title = "{Cooling of neutron stars with quark-hadron continuity}",
    booktitle = {European Physical Journal Web of Conferences},
         year = 2022,
       series = {European Physical Journal Web of Conferences},
       volume = {260},
        month = sep,
          eid = {11024},
        pages = {11024},
          doi = {10.1051/epjconf/202226011024},
       adsurl = {https://ui.adsabs.harvard.edu/abs/2022EPJWC.26011024N}
}

@ARTICLE{2021ApJ...918L..33R,
       author = {{Ritter}, Andreas and {Parker}, Quentin A. and {Lykou}, Foteini and {Zijlstra}, Albert A. and {Guerrero}, Mart{\'\i}n A. and {Le D{\^u}}, Pascal},
        title = "{The Remnant and Origin of the Historical Supernova 1181 AD}",
      journal = {\apjl},
         year = 2021,
        month = sep,
       volume = {918},
       number = {2},
          eid = {L33},
        pages = {L33},
          doi = {10.3847/2041-8213/ac2253},
archivePrefix = {arXiv},
       eprint = {2105.12384},
 primaryClass = {astro-ph.HE},
       adsurl = {https://ui.adsabs.harvard.edu/abs/2021ApJ...918L..33R}
}

@ARTICLE{2008RvMP...80.1455A,
       author = {{Alford}, Mark G. and {Schmitt}, Andreas and {Rajagopal}, Krishna and {Sch{\"a}fer}, Thomas},
        title = "{Color superconductivity in dense quark matter}",
      journal = {Reviews of Modern Physics},
         year = 2008,
        month = oct,
       volume = {80},
       number = {4},
        pages = {1455-1515},
          doi = {10.1103/RevModPhys.80.1455},
archivePrefix = {arXiv},
       eprint = {0709.4635},
 primaryClass = {hep-ph},
       adsurl = {https://ui.adsabs.harvard.edu/abs/2008RvMP...80.1455A}
}

@ARTICLE{2002JHEP...06..031A,
       author = {{Alford}, Mark and {Rajagopal}, Krishna},
        title = "{Absence of two-flavor color-superconductivity in compact stars}",
      journal = {Journal of High Energy Physics},
         year = 2002,
        month = jun,
       volume = {2002},
       number = {6},
          eid = {031},
        pages = {031},
          doi = {10.1088/1126-6708/2002/06/031},
archivePrefix = {arXiv},
       eprint = {hep-ph/0204001},
 primaryClass = {hep-ph},
       adsurl = {https://ui.adsabs.harvard.edu/abs/2002JHEP...06..031A}
}

@ARTICLE{2006NuPhA.768..118A,
       author = {{Abuki}, Hiroaki and {Kunihiro}, Teiji},
        title = "{Extensive study of phase diagram for charge-neutral homogeneous quark matter affected by dynamical chiral condensation: unified picture for thermal unpairing transitions from weak to strong coupling}",
      journal = {\nphysa},
         year = 2006,
        month = mar,
       volume = {768},
       number = {1},
        pages = {118-159},
          doi = {10.1016/j.nuclphysa.2005.12.019},
archivePrefix = {arXiv},
       eprint = {hep-ph/0509172},
 primaryClass = {hep-ph},
       adsurl = {https://ui.adsabs.harvard.edu/abs/2006NuPhA.768..118A}
}
\bibliographystyle{ptephy}
\end{document}